\documentclass[english,12pt,aps,prb,reprint]{revtex4-1}
\usepackage{lmodern}
\usepackage[T1]{fontenc}
\usepackage[latin9]{inputenc}
\usepackage{color}
\usepackage{babel}
\usepackage{array}
\usepackage{textcomp}
\usepackage{graphicx}
\usepackage{subscript}
\usepackage[unicode=true,pdfusetitle,
 bookmarks=true,bookmarksnumbered=false,bookmarksopen=false,
 breaklinks=false,pdfborder={0 0 1},backref=false,colorlinks=true]
 {hyperref}
\hypersetup{
 linkcolor=blue, citecolor=blue, urlcolor=blue, pdfstartview=XYZ, plainpages=false, pdfpagelabels}
\usepackage{breakurl}

\makeatletter

\providecommand{\tabularnewline}{\\}

\@ifundefined{date}{}{\date{}}

\makeatother

\begin{document}

\title{Delineating the role of ripples on thermal expansion of honeycomb
materials: graphene, 2D-h-BN and monolayer(ML)-MoS\textsubscript{2}}

\author{P. Anees}

\email{anees@igcar.gov.in}

\affiliation{\textsuperscript{}Materials Physics Division, Indira Gandhi Centre
for Atomic Research, HBNI, Kalpakkam 603 102, Tamil Nadu, India}

\author{M. C. Valsakumar}

\email{mc.valsakumar@gmail.com}

\affiliation{\textsuperscript{}Department of Physics, Indian Institute of Technology
Palakkad, Palakkad 678 557, Kerala, India}

\author{B. K. Panigrahi}

\affiliation{Materials Physics Division, Indira Gandhi Centre for Atomic Research,
HBNI, Kalpakkam 603 102, Tamil Nadu, India}

\email{bkp@igcar.gov.in}

\begin{abstract}
We delineated the role of thermally excited ripples on thermal expansion
properties of 2D honeycomb materials (free-standing graphene, 2D h-BN,
and ML-MoS\textsubscript{2}), by explicitly carrying out three-dimensional
(3D) and two-dimensional (2D) molecular dynamics simulations. In 3D
simulations, the in-plane lattice parameter (\textbf{\textit{a}}-lattice)
of graphene and 2D h-BN shows thermal contraction over a wide range
of temperatures and exhibits a strong system size dependence. The
2D simulations of the very same system show a reverse trend, where
the \textbf{\textit{a}}-lattice is expanding in the whole computed
temperature range. Contrary to graphene and 2D h-BN, the \textbf{\textit{a}}-lattice
of ML-MoS\textsubscript{2} shows thermal expansion in both 2D and
3D simulations and their system size dependence is marginal. By analyzing
the phonon dispersion at 300 K, we found that the discrepancy between
2D and 3D simulations of graphene and 2D h-BN is due to the absence
of out-of-plane bending mode (ZA) in 2D simulations, which is responsible
for thermal contraction of \textbf{\textit{a}}-lattice at low temperature.
Meanwhile, all the phonon modes are present in 2D phonon dispersion
of ML-MoS\textsubscript{2}, which indicates that the origin of ZA
mode is not purely due to out-of-plane movement of atoms and also
its effect on thermal expansion is not significant as found in graphene
and 2D h-BN. 
\end{abstract}
\maketitle

\section{Introduction}

Graphene has got enormous attraction due its fascinating electronic,
thermal and mechanical properties\cite{geim2007rise,Balandin2011,Lee18072008},
and it is proposed as a promising candidate for next generation electronic
industry\cite{geim2007rise,Schwierz2010,RevModPhys.81.109}. The major
pitfall in graphene based electronics is the absence of finite band
gap in its electronic band structure. After successful isolation of
graphene, the search for other 2D honeycomb materials were geared
up in past few years. The 2D hexagonal (h)-BN, which is isostructural
to graphene, is an insulator with a finite band gap \textasciitilde{}5-6
eV \cite{doi:10.1021/nn1006495}, and exhibits intriguing electronic
properties\cite{doi:10.1021/nn1006495,doi:10.1021/nl1023707,Watanabe2004}.
The family of 2D materials are getting richer day by day\cite{Geim2013}.
Apart from graphene and h-BN, monolayer(ML)-MoS\textsubscript{2}
is another high interesting 2D honeycomb material\cite{Chhowalla2013,C2CS35387C}.
ML-MoS\textsubscript{2} is a direct band gap\cite{Chhowalla2013}
(1.9 eV) semiconductor, and it exhibits high photoluminescence yield\cite{doi:10.1021/nl903868w},
which puts this material in the front-end of optoelectronic industry.
The ML-MoS\textsubscript{2} based field effect transistors (FETs)
shows high carrier mobility\cite{RadisavljevicB2011} and on/off ratios\cite{RadisavljevicB2011,C2NR33443G,C3NR04218A}. 

Structural stability of the 2D crystals was an old dispute in condensed
matter theory. According to Mermin-Wagner theorem\cite{PhysRev.176.250},
the long wavelength thermal fluctuations will destroy the long-range
order in 2D crystals. But in the case of graphene and 2D h-BN, these
fluctuations are suppressed by strong anharmonic coupling between
in-plane stretching and out-of-plane bending modes, leading to height
fluctuations on the surface, known as ripples\cite{Fasolino2007}.
These intrinsic ripples are inevitable in 2D crystals, and they stabilizes
the 2D membranes \cite{Fasolino2007,2053-1583-2-3-035014,C5CP06111C}.
Transmission electron microscopic study reveals that, the suspended
graphene sheets are not perfectly flat, they exhibits out-of-plane
deformations\cite{Meyer2007}. Recent experiments using high resolution
atomic force microscopy, shows sinusoidal ripples of periodicity 3
to 6 nm and amplitude of 10 to 100 pm on the surfaces of graphene
and 2D h-BN layer of supported flakes\cite{2015arXiv150405253G}.
The ripples structure in graphene could be manipulated to sketch devices
based on local strain\cite{2008arXiv0810.4539P} and band gap engineering\cite{Elias610}. 

For aforementioned applications of 2D materials, knowledge of linear
thermal expansion coefficients (LTECs) is essential. Several studies
has been reported the LTEC of graphene both from simulations\cite{PhysRevB.71.205214,PhysRevLett.102.046808,PhysRevLett.106.135501,PhysRevB.89.035422,2053-1583-2-3-035014}
and experiments\cite{Bao2009,0957-4484-21-16-165204,doi:10.1021/nl201488g,Pan2012}.
Mounet and Marzari \textit{\cite{PhysRevB.71.205214}} predicts that
the LTECs of graphene remains to be negative upto 2300 K using quasi-harmonic
calculations. Zakhrchenko\textit{ et al\cite{PhysRevLett.102.046808}
}performed Monte-Carlo (MC) simulations and found that in-plane lattice
parameter (\textbf{\textit{a}}-lattice) contracts with temperature
upto T = 900 K and further it expands. \textit{Ab initio} molecular
dynamics (MD) simulations by Pozzo \textit{et al\cite{PhysRevLett.106.135501}
}shows that the C-C distance increases with an increase in temperature
for both supported and free-standing graphene; meanwhile the \textbf{\textit{a}}-lattice
is found to be contracting with an increase in temperature (upto 2000
K) in free-standing graphene. The above discrepancies among the various
simulations arises due to the difference in the incorporation of anharmonicity
in those calculations, its effects are very strong in 2D crystals\cite{2053-1583-2-3-035014}.
From the experimental front, Bao \textit{et al\cite{Bao2009}}, reported
the negative thermal expansion of graphene in the temperature range
200 K - 400 K. Later, Yoon \textit{et al\cite{doi:10.1021/nl201488g},}
also found that thermal expansion coefficient of graphene is negative
in the above temperature range using temperature dependent Raman spectroscopy.
The authors also observed the strain effect induced by substrate-layer
interaction can alter the physical properties of graphene. 

In quasi-harmonic approximation (QHA) the 2D sheets are considered
to be flat, hence the effects of ripples cannot be incorporated in
a direct manner. Moreover, within the \textit{ab initio} frame work
we cannot include more than few hundreds of atoms in the simulation
cell, which seems to be in-adequate to incorporate the long wavelength
ripples. From the experimental perspective, most of the measurements
are made on graphene supported on a substrate or over a trench, such
measurements are extremely challenging due to the strain effects,
and may not be able to capture the intrinsic thermal expansion properties
of free-standing graphene with ripples. Classical MD simulations can
incorporate millions of atoms and also computation can be done with
free-standing sheets contains all rippling effects, hence it will
be an ideal choice to overcome the above limitations. The thermal
expansion of graphene has been reported in various studies as mentioned
above. However, the thermal expansion of 2D h-BN and ML-MoS\textsubscript{2}
are not studied in detail, which is essential to devise hybrid nano-devices
and hetero-structures \cite{Geim2013}. The objective of the present
paper is to understand the the role of ripples on thermal expansion
properties of honeycomb materials explicitly. To delineate the role
of ripples, we studied the the thermal expansions of very same system
using three-dimensional (3D) and two-dimensional (2D) molecular dynamics
(MD) simulations, the later cannot incorporate the effects ripples.

\section{Computational methods\label{sec:Computational-methods}}

All simulations are done using classical MD simulation package LAMMPS\cite{Plimpton19951}.
To understand the role of ripples, we explicitly carried out 2D and
3D simulations of very same system at different temperatures. In 2D
simulations, we arrested the motion of atom along the direction normal
to the sheet and prevent the formation of thermally excited ripples
using \textit{fix enforce2d }command\cite{Plimpton19951}. Ripples
are naturally included in 3D simulations and leads to a corrugated
surface instead of flat 2D sheet. Simulation cell of different sizes
are used to incorporate the effects of long wavelength ripples. Periodic
boundary conditions (PBC) are employed in all the three directions.
To avoid the un-physical interactions between the periodic images,
the sheets are stacked one above another with an additional vacuum
separation of 15 Å. In this study, we employed empirical interatomic
potentials (EIP) to model the interactions in honeycomb structures.
We followed the same algorithm for both 2D and 3D simulations which
is given below. Inorder to eliminate any residual stresses that could
be present in the initial configuration, the geometry is relaxed using
conjugate gradient algorithm. The system is then equilibrated for
500 picoseconds (ps) in isobaric-isothermal (NPT) ensemble at desired
temperatures and ambient pressure. After ensuring the proper equilibration
and thermalization, we monitored the variation of lattice parameters
as a function temperatures. The whole simulations are done for 3.2
nanoseconds. The linear thermal expansion coefficients (LTECs) are
obtained by direct numerical differentiation (equation \ref{eq:1})
of above data.

\begin{equation}
\alpha\left(T\right)=\frac{1}{a\left(T\right)}\frac{da\left(T\right)}{dT}\label{eq:1}
\end{equation}

\section{Results and Discussions}

\begin{figure*}
\begin{centering}
\includegraphics[scale=0.497]{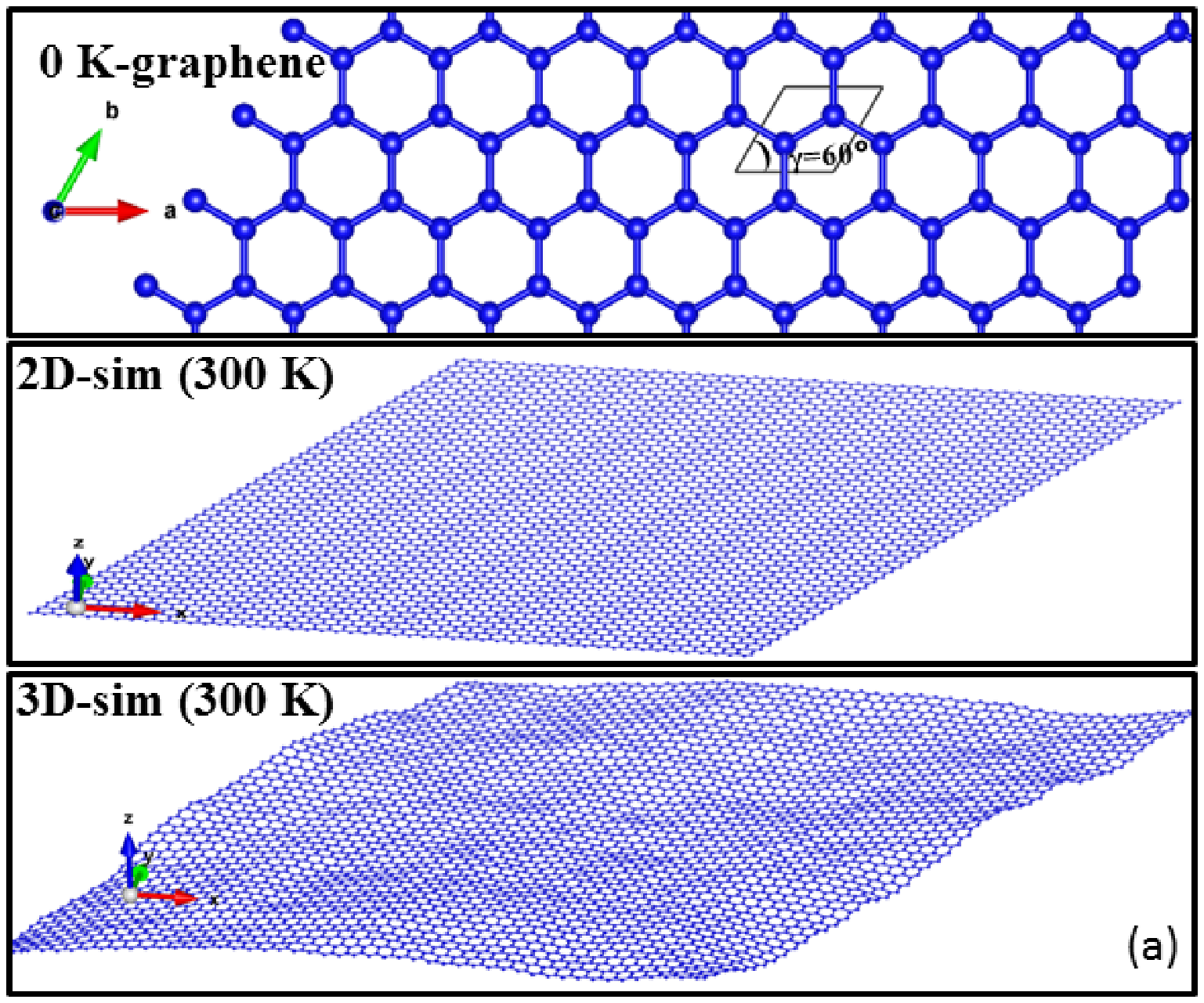}\includegraphics[scale=0.503]{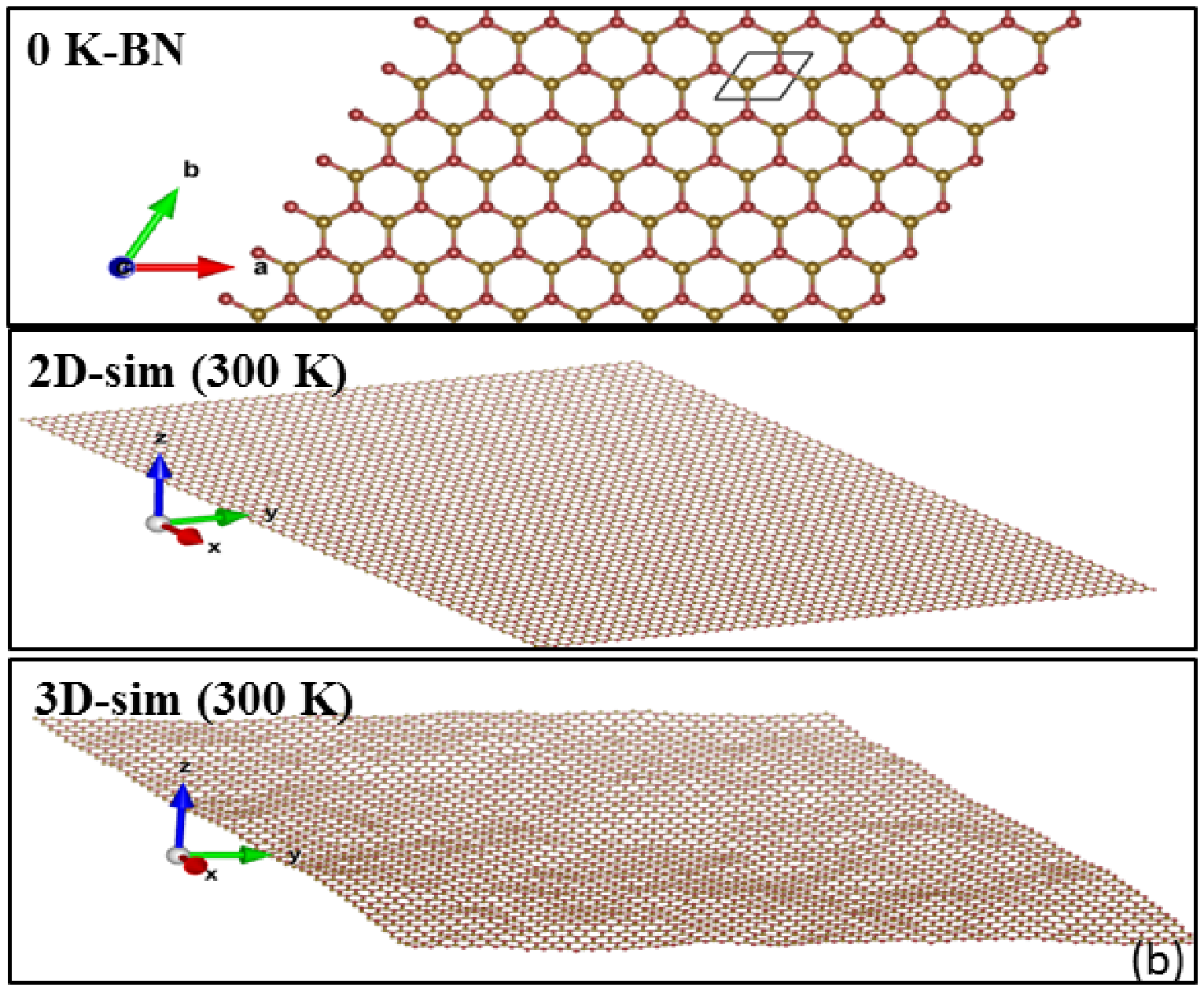}
\par\end{centering}

\begin{centering}
\includegraphics[scale=0.5]{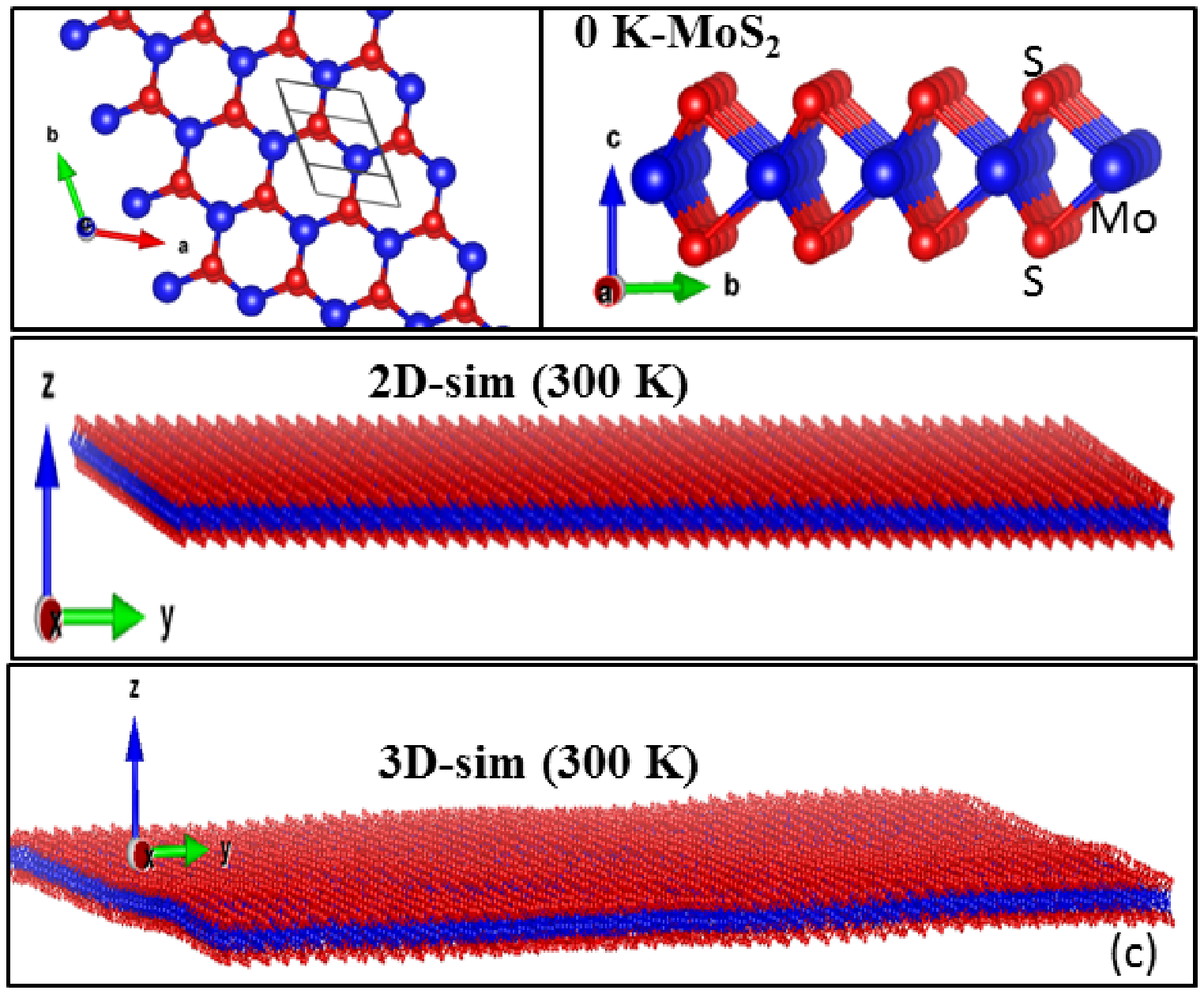}
\par\end{centering}

\protect\caption{\label{fig:Fig1-structres}The 2D honeycomb lattices of (a) graphene,
(b) 2D h-BN and (c) ML-MoS\protect\textsubscript{2}. (top) The 2D
sheets at 0 K, the unitcells can be represented using a rhombus, and
the corresponding primitive translation vectors are $\protect\overrightarrow{a}$=(a,0,0),
$\protect\overrightarrow{b}$ =(a/2,$\sqrt{3}$ a/2,0) and $\protect\overrightarrow{c}$
=(0,0,c); (middle) snapshots of sheets obtained from 2D simulations
at 300 K. (bottom) sheets obtained from 3D simulation at 300 K, and
they are no longer flat as in 2D simulations, the height fluctuation
(ripples) normal to the surfaces are conspicuous.}
\end{figure*}

\begin{figure*}
\begin{centering}
\includegraphics[scale=0.38]{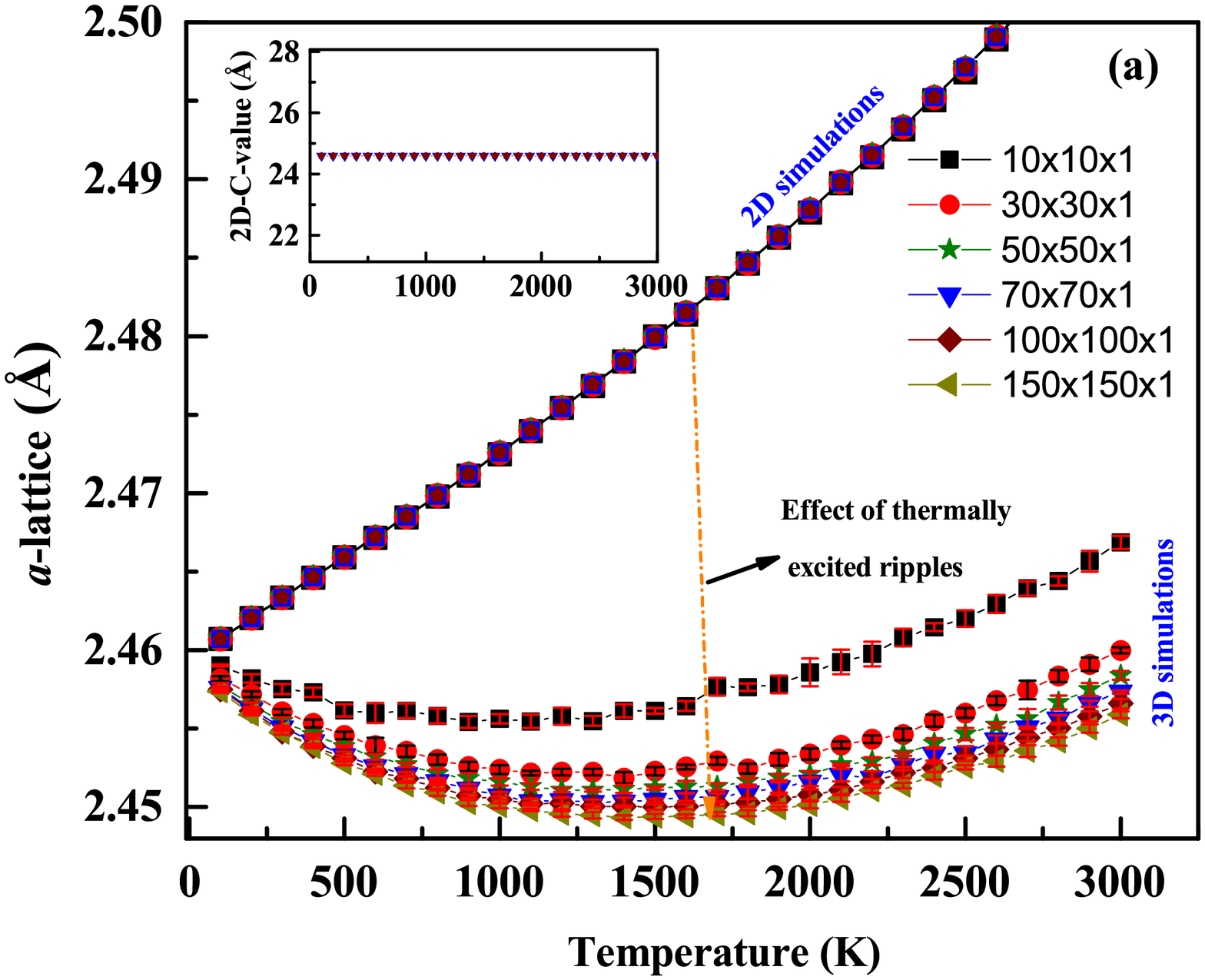}\includegraphics[scale=0.38]{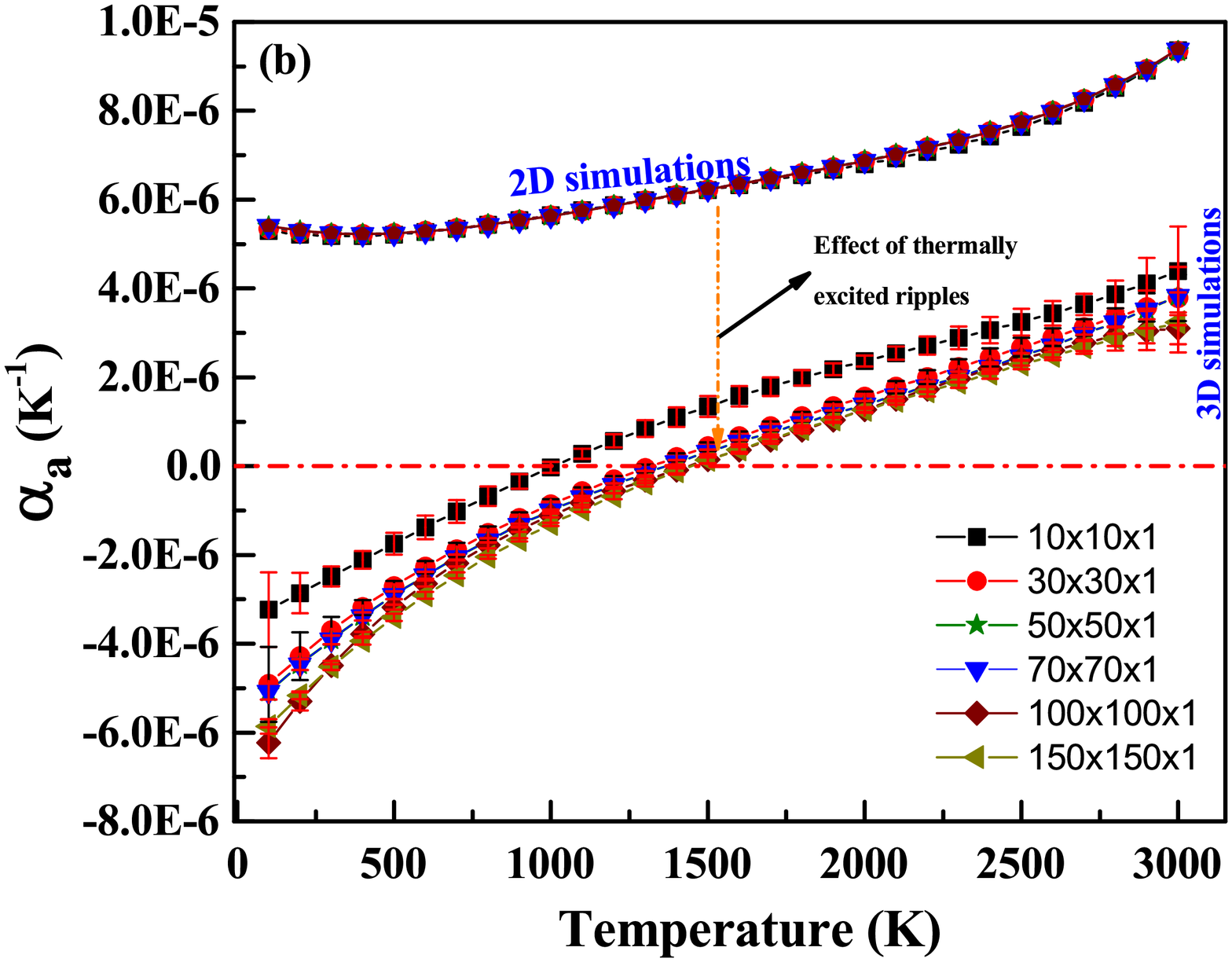}
\par\end{centering}

\protect\caption{\label{fig:The-graphene-thermal-expn}(a) The temperature evolution
of in-plane lattice parameter (\textbf{\textit{a}}-lattice) of graphene
obtained from 2D and 3D simulations. (Inset) C-value obtained from
2D simulations, which doesn't changes with temperature. (b) The linear
thermal coefficients (LTECs) as a function of temperature. In 3D simulations,
the LTEC changes its sign from negative to positive in the temperature
range of 1100 K - 1400 K (depends on system size).}
\end{figure*}

Figure \ref{fig:Fig1-structres} displays the configuration of graphene,
2D h-BN and ML-MoS\textsubscript{2} at 300 K. In 2D simulation, we
obtained a flat 2D sheet without any ripples, while 3D simulation
shows corrugations on the surface due to the formation of ripples.
The interactions between C-C atoms in honeycomb lattice of graphene
is modeled using a bond-order potential (LCBOP)\cite{PhysRevB.68.024107}.
The LCBOP potential predict the equilibrium in-plane lattice parameter
(\textbf{\textit{a}}-lattice) \textbf{\textit{a\textsubscript{\textbf{\textit{0}}}=
}}2.459 Å, shows an excellent agreement with experiment (\textbf{\textit{a\textsubscript{\textbf{\textit{0}}}=
}}2.463 Å). We used simulation cells of various sizes (10\texttimes 10\texttimes 1,
30\texttimes 30\texttimes 1, 50\texttimes 50\texttimes 1, 70\texttimes 70\texttimes 1,
100\texttimes 100\texttimes 1, 150\texttimes 150\texttimes 1) to incorporate
the effect of long wavelength ripples. Figure \ref{fig:The-graphene-thermal-expn}
displays the temperature dependence of \textbf{\textit{a}}-lattice
and linear LTEC of free-standing graphene. In 3D simulations, we found
that the \textbf{\textit{a}}-lattice decreases with an increase in
temperature. Fourth order polynomial fit to the above data shows that
minima occurs in the temperature range 1300 K - 1400 K, and further
it expands with an increase in temperature, this is consistent with
our previous study\cite{2053-1583-2-3-035014}. 

Noteworthy, the temperature evolution of \textbf{\textit{a}}-lattice
is system size dependent (Figure \ref{fig:Fig1-structres}). For simulation
cell of size 10\texttimes 10\texttimes 1 (contains only 200 atoms)
\textbf{\textit{a}}-lattice shows relatively less contraction with
respect to bigger cells and minima occurs around T = 1100 K. As we
increase the system size size, \textbf{\textit{a}}-lattice shows a
convergence from 70\texttimes 70\texttimes 1 (9800 atoms) onwards,
and the minima falls in the temperature range 1300 K - 1400 K (dependence
on system size). The similar system size dependence was reported by
Pozzo \textit{et al\cite{PhysRevLett.106.135501}}, where they used
simulation cells of sizes 8\texttimes 8\texttimes 1, 10\texttimes 10\texttimes 1
and 16\texttimes 16\texttimes 1 contains 128, 200 and 512 atoms, respectively.
Fasolino\textit{ et al\cite{Fasolino2007}} observed fluctuations
with wavelength of the order of 80 Å at 300 K from their Monte Carlo
simulations. To incorporate such long wave length fluctuations bigger
simulation cells are required, which makes\textit{ ab initio} Car-Parinello
simulations prohibitive. In the present study, we used a simulation
cell of size 150\texttimes 150\texttimes 1 (45000 atoms), which is
capable of incorporating all long-wavelength rippling effects.

\begin{table*}
\centering{}%
\begin{tabular}{>{\raggedright}p{0.24\textwidth}>{\centering}p{0.12\textwidth}>{\centering}p{0.12\textwidth}>{\centering}p{0.22\textwidth}}
\hline 
\multicolumn{1}{>{\raggedright}p{0.1\textwidth}}{simulation cell size } & 2D simulation 

$\alpha_{a}(\times10^{-6}K^{-1})$ & 3D simulation 

$\alpha_{a}(\times10^{-6}K^{-1})$ & expt.

$\alpha_{a}(\times10^{-6}K^{-1})$\tabularnewline
\hline 
\multicolumn{1}{l}{10\texttimes 10\texttimes 1 (200 atoms)} & \multicolumn{1}{c}{5.178} & -2.499 & \tabularnewline
30\texttimes 30\texttimes 1 (1800 atoms) & 5.226 & -4.095 & \tabularnewline
50\texttimes 50\texttimes 1 (5000 atoms) & 5.235 & -4.100 & -5.500\textsuperscript{a}, -7.000\textsuperscript{b,c}, -8.000\textsuperscript{d}\tabularnewline
70\texttimes 70\texttimes 1 (9800 atoms) & 5.230 & -4.380 & \tabularnewline
100\texttimes 100\texttimes 1 (20000 atoms) & 5.243 & -4.350 & \tabularnewline
150\texttimes 150\texttimes 1 (45000 atoms) & 5.241 & -4.524 & \tabularnewline
\hline 
\end{tabular}\protect\caption{\label{tab:The-LTECs-at-300K-graphene}The system size dependence
of linear thermal expansion coefficients (LTECs) of graphene at 300
K. In 2D simulations, the LTECs are positive and does not show any
system size dependence. The LTECs obtained from 3D simulations all
are negative and shows a system size dependence. The data has been
compared with the experiments. \protect\textsuperscript{a}Reference\cite{Pan2012},
\protect\textsuperscript{b}Reference\cite{Bao2009}, \protect\textsuperscript{c}Reference\cite{0957-4484-21-16-165204},
\protect\textsuperscript{d}Reference \cite{doi:10.1021/nl201488g}. }
\end{table*}

In 3D simulations, the LTEC are negative for all simulation cells.
The LTEC also shows a system size dependence and its value for 10\texttimes 10\texttimes 1
simulation cell is roughly half of the value of 150\texttimes 150\texttimes 1
cell (table \ref{tab:The-LTECs-at-300K-graphene}). The value of LTEC
at 300 K $(\alpha_{a}=-4.35\times10^{-6}K^{-1})$ is in qualitative
agreement with previous calculations\cite{PhysRevB.71.205214,PhysRevLett.102.046808,PhysRevB.89.035422}.
All simulations predict the LTEC roughly half of the experimental
value\cite{Bao2009,doi:10.1021/nl201488g} (table \ref{tab:The-LTECs-at-300K-graphene}).
Unlike DFPT calculations, present study incorporated the full anharmonicity
of interatomic potential, hence the disagreement with experimental
data may not be due to the strong anharmonic nature of graphene. Though
the above experiments \cite{Bao2009,doi:10.1021/nl201488g} has taken
care to eliminate the strain effect induced by substrate, more accurate
analysis is needed to get a clear picture. To support the above arguments,
we can see the previous observation of Pozzo \textit{et al\cite{PhysRevLett.106.135501}},
when the graphene sheet was supported on Ir (111) substrate, it shows
thermal expansion instead of thermal contraction. Jiang \textit{et
al\cite{PhysRevB.80.205429} }used Green's function technique and
reported that the LTEC is very sensitive to substrate layer interaction,
a weak substrate-layer interaction can cause a significant change
in the value of LTEC, and if the substrate effects are strong enough,
the LTEC can become positive in the whole computed temperature range.
Later, Pan \textit{et al\cite{Pan2012}, }used temperature dependent
Raman spectroscopy and\textit{ }measured a lower bound of LTEC (at
300 K $(\alpha_{a}=-5.5\times10^{-6}K^{-1})$ of graphene which was
supported on BN, while Bao \textit{et al\cite{Bao2009}} and Yoon\textit{
et al\cite{doi:10.1021/nl201488g}} used Si and SiO\textsubscript{2}
substrates to support their graphene sheet, respectively; this may
be one of the reason to have different LTEC in these experiments.
Our results, along with the earlier theoretical predictions\cite{PhysRevB.71.205214,PhysRevLett.102.046808,PhysRevB.89.035422}
are in qualitative agreement with Pan \textit{et al\cite{Pan2012}. }

The temperature dependence of \textbf{\textit{a}}-lattice obtained
from 2D simulations is shown in Figure \ref{fig:Fig1-structres}.
In contrast to 3D simulations, \textbf{\textit{a}}-lattice increases
with an increase in temperature, and it does not show any system size
dependence.\textit{ }Unlike 3D simulations, the LTEC obtained from
2D simulations are all positive in sign and does not have any system
size dependence (table \ref{tab:The-LTECs-at-300K-graphene}). Since
there is no movement of atoms along Z direction, rippling effects
are absent in 2D simulations, hence \textbf{\textit{a}}-lattice shows
a thermal expansion, and the sign of LTEC is positive in the whole
computed temperature range. From above observations, it can be concluded
that, the long wavelength ripples are responsible for thermal contraction
of free-standing graphene over a wide range of temperatures.

\begin{figure*}
\begin{centering}
\includegraphics[scale=0.38]{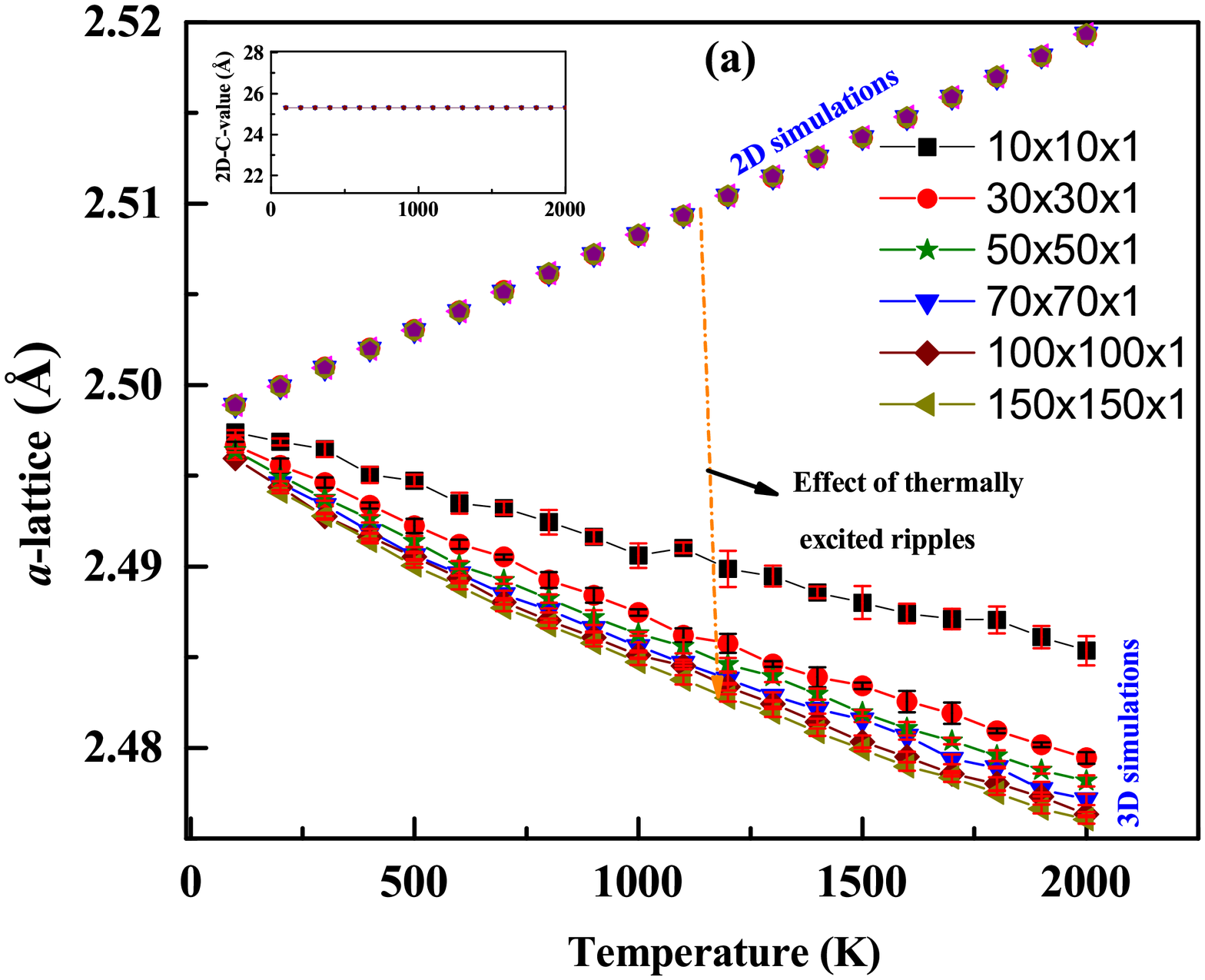}\includegraphics[scale=0.38]{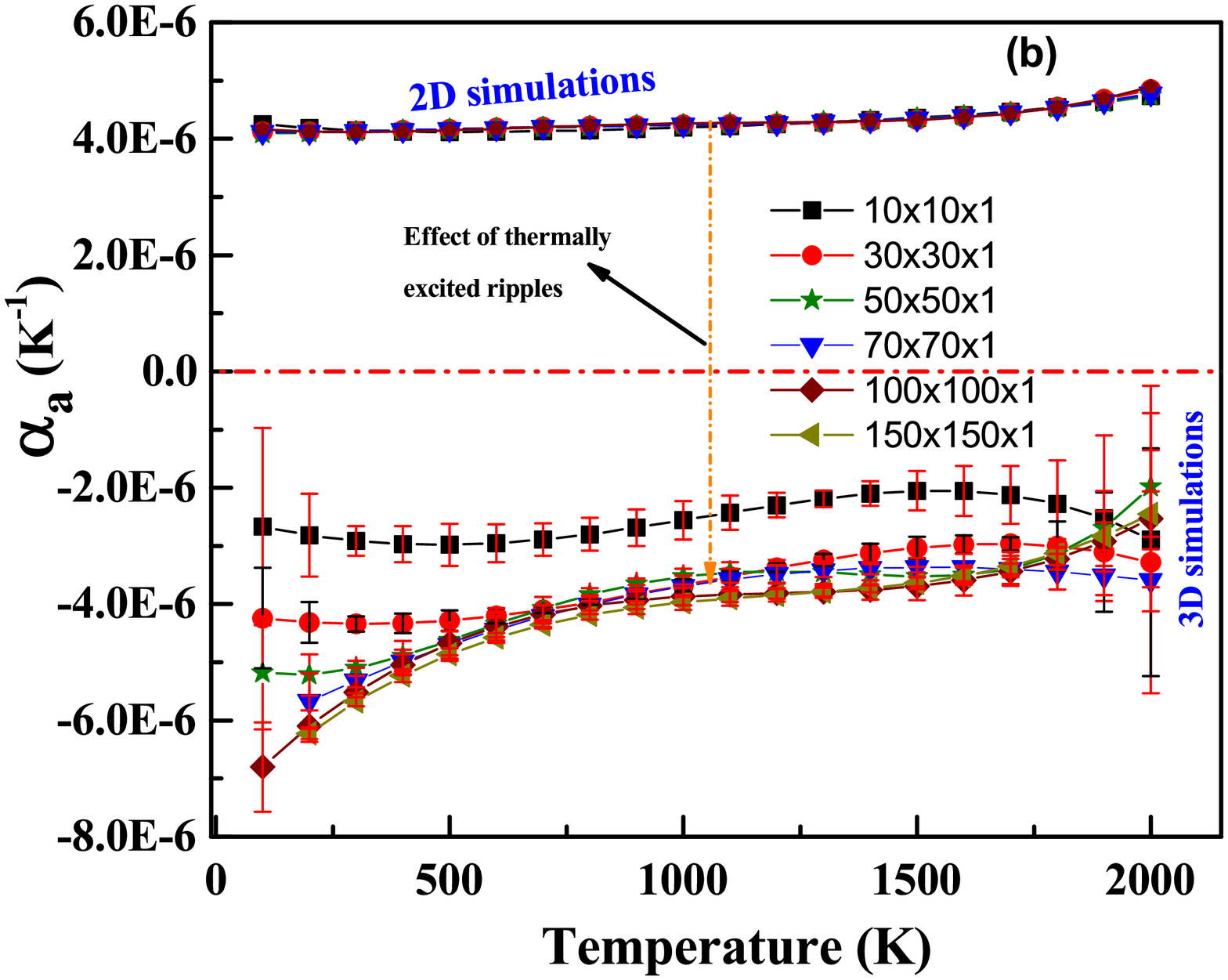}
\par\end{centering}

\protect\caption{\label{fig:the-h-BN-thermal-expn} (a) The temperature dependence
of in-plane lattice parameter of (\textbf{\textit{a}}-lattice) of
2D h-BN; (b) Linear thermal expansion coefficients (LTECs) as a function
of temperature}
\end{figure*}

2D h-BN is another one-atom thick material, being 2D crystal, ripples
are un-avoidable in 2D h-BN also; hence we extended the above analysis
to 2D h-BN to understand its thermal expansion behaviour. The interaction
between the B and N atom is modeled using a Tersoff type potential
parametrized by Sevik \textit{et al}\cite{PhysRevB.84.085409}. The
present potential predicts the structural and mechanical properties
of 2D h-BN reasonably accurately. The equilibrium \textbf{\textit{a}}-lattice
obtained with present potential shows an excellent matching with experiments
(\textbf{\textit{a\textsubscript{\textbf{\textit{0}}}= }}2.500 Å)\cite{Pacile2008,doi:10.1021/nn9018762}.
Figure \ref{fig:the-h-BN-thermal-expn} displays the temperature dependence
of \textbf{\textit{a}}-lattice. In 3D simulation, \textbf{\textit{a}}-lattice
decreases with an increase in temperature in the whole computed range
and matching with our previous study \cite{C5CP06111C}. Similar to
graphene, \textbf{\textit{a}}-lattice shows a system size dependence
in 2D h-BN also, and again we found a convergence from the simulation
cell of size 70\texttimes 70\texttimes 1 (9800) onwards. Paszkowicz\textit{
et al\cite{raey}} measured the thermal expansion (10 K - 297.5 K)
of bulk h-BN using synchrotron X-ray diffraction technique, and they
found that \textbf{\textit{a}}-lattice shows a flat variation at low
temperatures (10 K - 100 K), above 100 K it falls with an increase
in temperatures upto 300 K. In 2D simulation, the \textbf{\textit{a}}-lattice
increases with an increase in temperature and shows thermal expansion
in the whole computed temperature range. The \textbf{\textit{a}}-lattice
does not show any system size dependence in 2D simulations and it
is consistent with our observations in graphene. The LTEC obtained
at 300 K are shown in table \ref{tab:The-LTECs-at-300 K-BN}, the
system size dependence of LTEC is discernible. The LTEC at 300 K $(\alpha_{a}=-5.509\times10^{-6}K^{-1})$
is matching with previous quasi-harmonic predictions\cite{PhysRevB.89.035422}.
The LTEC obtained from 2D simulations all are positive in sign and
does not have any system size dependence. Like graphene, the effect
of ripples are quite strong in 2D h-BN also, since the empirical potential
used to study 2D h-BN and graphene are different, we are not attempting
a direct comparison among their data.

\begin{table*}
\centering{}%
\begin{tabular}{>{\raggedright}p{0.22\textwidth}>{\centering}p{0.15\textwidth}>{\centering}p{0.15\textwidth}}
\hline 
\multicolumn{1}{>{\raggedright}p{0.1\textwidth}}{simulation cell size } & 2D simulation 

$\alpha_{a}(\times10^{-6}K^{-1})$ & 3D simulation 

$\alpha_{a}(\times10^{-6}K^{-1})$\tabularnewline
\hline 
\multicolumn{1}{l}{10\texttimes 10\texttimes 1 (200 atoms)} & \multicolumn{1}{c}{4.145} & -2.940\tabularnewline
30\texttimes 30\texttimes 1 (1800 atoms) & 4.130 & -4.272\tabularnewline
50\texttimes 50\texttimes 1 (5000 atoms) & 4.124 & -5.151\tabularnewline
70\texttimes 70\texttimes 1 (9800 atoms) & 4.129 & -5.508\tabularnewline
100\texttimes 100\texttimes 1 (20000 atoms) & 4.114 & -5.509\tabularnewline
150\texttimes 150\texttimes 1 (45000 atoms) & 4.107 & -5.670\tabularnewline
\hline 
\end{tabular}\protect\caption{\label{tab:The-LTECs-at-300 K-BN}The linear thermal expansion coefficients
(LTECs) of 2D h-BN at 300 K, the system size dependence of LTECs obtained
from 3D simulations are discernible. }
\end{table*}

\begin{figure*}
\centering{}\includegraphics[scale=0.4]{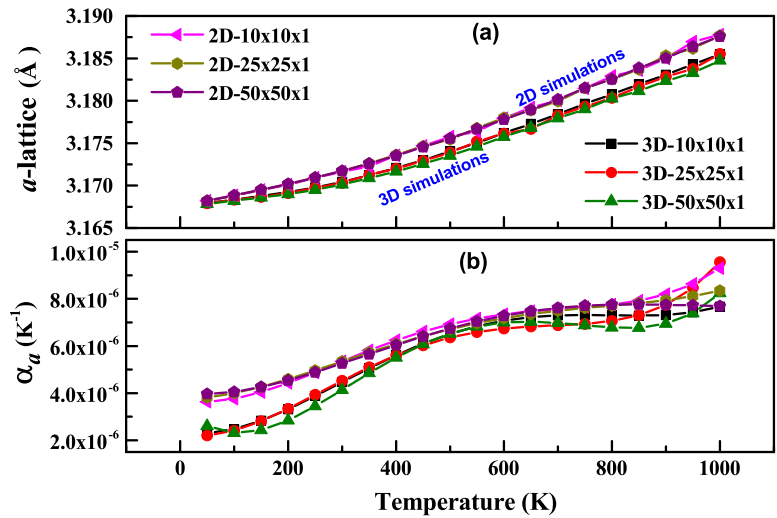}\protect\caption{\label{fig:The-MoS2-therm-expn} (a) Variation of in-plane lattice
parameter (\textbf{\textit{a}}-lattice) of ML-MoS\protect\textsubscript{2}
with temperature; (b) The linear thermal expansion coefficients (LTECs)
as a function of temperature. Unlike graphene and 2D-h-BN, the system
size dependence of \textbf{\textit{a}}-lattice is marginal in ML-MoS\protect\textsubscript{2}.}
\end{figure*}

Apart from graphene and 2D-h-BN, ML-MoS\textsubscript{2} is another
high interesting honeycomb material. The Mo atom layer sandwiched
in between two S atom layers (S-Mo-S sandwich structures) in a trigonal
prismatic fashion (Figure \ref{fig:Fig1-structres}). Liang \textit{et
al\cite{PhysRevB.79.245110}} parametrized a many body reactive empirical
bond order (REBO) potential for Mo-S system. This potential could
successfully model the structural and mechanical properties of Mo-S
and MoS\textsubscript{2} systems. Later, Stewart and Spearot \textit{\cite{0965-0393-21-4-045003}
}refined the parametrization and implemented it into MD simulation
package LAMMPS. We used the parametrization of Stewart\textit{ }and
Spearot\textit{\cite{0965-0393-21-4-045003}} to model the interaction
between the Mo and S atoms in ML-MoS\textsubscript{2}. The present
potential predicts the equilibrium \textbf{\textit{a}}-lattice,\textbf{\textit{
a\textsubscript{\textbf{\textit{0}}}}} = 3.17 Å which is close to
the experimentally reported value (3.16 Å)\cite{doi:10.1021/jp205116x}.
Figure \ref{fig:The-MoS2-therm-expn} displays the thermal expansion
of \textbf{\textit{a}}-lattice and LTEC. The \textbf{\textit{a}}-lattice
is expanding in the whole computed temperature range in both 2D and
3D simulations, and its system size dependence is marginal. The LTECs
of ML-MoS\textsubscript{2} is positive in both 2D and 3D simulations
(table \ref{tab:The-LTECs-at-300K-MoS2}), and their magnitudes are
slightly higher in 2D simulations at low temperatures (T < 300 K).
The LTEC obtained from 3D simulation at 300 K (4.140$\times10^{-6}K^{-1}$)
matches with previous experimental data (4.922$\times10^{-6}K^{-1})$\cite{JCR:JCRA18163}.
Though ML-MoS\textsubscript{2} possess the same hexagonal honeycomb
lattice structure of graphene and 2D h-BN, the \textbf{\textit{a}}-lattice
shows a positive thermal expansion\cite{PhysRevB.90.045409,PhysRevB.89.035422,:/content/aip/journal/apl/108/10/10.1063/1.4943546}
and it has been measured earlier in bulk-MoS\textsubscript{2} using
X-ray powder diffraction\cite{JCR:JCRA13982,JCR:JCRA18163}. This
contradiction with graphene and 2D h-BN can be visualized as an effect
of S-Mo-S sandwich structure in ML-MoS\textsubscript{2}, which reduces
the rippling behavior considerably\cite{PhysRevB.91.014101}. The
Mo-Mo distance in Mo layer is higher than that of C-C atom in graphene,
hence the Mo-S interaction is responsible for lower height fluctuation
of Mo atom in ML-MoS\textsubscript{2}\cite{PhysRevB.91.014101}.
This difference in thermal expansion property among the above mentioned
honeycomb materials can be utilized to make future hybrid nano-devices. 

\begin{table*}
\centering{}%
\begin{tabular}{>{\raggedright}p{0.22\textwidth}>{\centering}p{0.15\textwidth}>{\centering}p{0.15\textwidth}}
\hline 
\multicolumn{1}{>{\raggedright}p{0.1\textwidth}}{simulation cell size } & 2D simulation 

$\alpha_{a}(\times10^{-6}K^{-1})$ & 3D simulation 

$\alpha_{a}(\times10^{-6}K^{-1})$\tabularnewline
\hline 
\multicolumn{1}{l}{10\texttimes 10\texttimes 1 (300)} & \multicolumn{1}{c}{5.340} & 4.469\tabularnewline
25\texttimes 25\texttimes 1 (1875 atoms) & 5.343 & 4.530\tabularnewline
50\texttimes 50\texttimes 1 (7500 atoms) & 5.261 & 4.140\tabularnewline
\hline 
\end{tabular}\protect\caption{\label{tab:The-LTECs-at-300K-MoS2}The LTECs of ML-MoS\protect\textsubscript{2 }at
300 K. The system size dependence of LTECs in both 3D and 2D simulations
are insignificant. }
\end{table*}

Inorder to understand the underlying mechanism behind the thermal
contraction or expansion of solids, Grüneisen theory has been widely
used\cite{PhysRevB.68.035425,PhysRevB.71.205214,PhysRevB.89.035422}.
According to Grüneisen theory, modes with positive Grüneisen parameters
will encourage the thermal expansion, while modes with negative Grüneisen
parameter will aid thermal contraction. A solid will undergo thermal
expansion or contraction is determined by the balance between the
modes with positive and negative Grüneisen parameters\cite{PhysRevB.68.035425}.
For graphene, the Grüneisen parameters of low lying bending mode (ZA)
become large negative (as low as -80).  At low temperature only low
frequency acoustic modes will be excited, (high frequency optic modes
with positive Grüneisen parameters are frozen) and contributes to
thermal contraction\cite{PhysRevB.71.205214,PhysRevB.89.035422}.
The negative Grüneisen parameter of ZA mode is due to the membrane
effect, predicted by Lifshitz\cite{Lifshitz1952}. Apart from graphene,
Sevik \textit{et al\cite{PhysRevB.89.035422}} extended Grüneisen
theory analysis to 2D h-BN and ML-MoS\textsubscript{2}, and they
observed a large negative Grüneisen parameter of ZA modes in 2D h-BN
also, and it is responsible for thermal contraction. For ML-MoS\textsubscript{2},
the Grüneisen parameter associated with ZA mode is relatively small
(\textasciitilde{} -10), and leads to thermal contraction only at
very low temperature (T < 20 K)\cite{PhysRevB.90.045409}. MD simulations
are not meaningful at very low temperatures, due to the manifestation
of quantum effects. Hence, we performed simulations for temperatures
T >100 K, so we couldn't observe above thermal contraction effects
in ML-MoS\textsubscript{2}. The Grüneisen parameter of ZA is fully
negative in graphene\cite{PhysRevB.71.205214}, while in ML-MoS\textsubscript{2}
it is negative only near the $\Gamma$ point and becomes positive
along K-M direction in the Brillouin zone\cite{PhysRevB.90.045409}
. This negative-to-positive change attributes to phononic hybridization
and finite thickness effects of ML-MoS\textsubscript{2} which counteracts
the membrane effects in 2D systems\cite{PhysRevB.90.045409}. 

The mode dependent Grüneisen parameters are computed by strain derivative
of phonon frequencies which obtained using quasi-harmonic approximation
(QHA)\cite{PhysRevB.71.205214}. One drawback of above method is that,
under certain compressive strain, it is difficult to keep the crystal
stable. When compressive strain is large enough, it leads to imaginary
frequencies around the $\Gamma$ point which cannot be used to compute
the mode Grüneisen parameters. Due to above limitation, wavevectors
are computed with less accurate finite difference algorithm around
the $\Gamma$ point\cite{Karssemeijer2010}. Moreover in QHA we are
using a flat 2D sheet, which is devoid of ripples. Despite the above
limitations, Grüneisen theory predicts the thermal expansion of honeycomb
structures reasonably well . 

Instead of Grüneisen theory, here we analyzed the role of different
phonon modes on thermal expansion behavior by computing the phonon
dispersion at finite temperatures. Inorder to understand the effect
of ripples on phonon modes, one has to compute the phonon dispersion
separately from 2D and 3D simulations. Lattice dynamics (LD) methods\cite{Dove1993}
predicts the phonon frequencies and polarizations from the second
derivative of interatomic potential at 0 K, hence it won't make any
difference between 2D and 3D simulations, moreover anharmonic effects
are completely absent in LD methods. To over come the above limitation,
we developed a spectral energy density based method to compute the
phonon frequencies directly from classical MD simulations \cite{2053-1583-2-3-035014,C5CP06111C},
it will capture the true anharmonic behavior of all phonon modes without
any approximations. 

\begin{figure*}
\centering{}\includegraphics[scale=0.33]{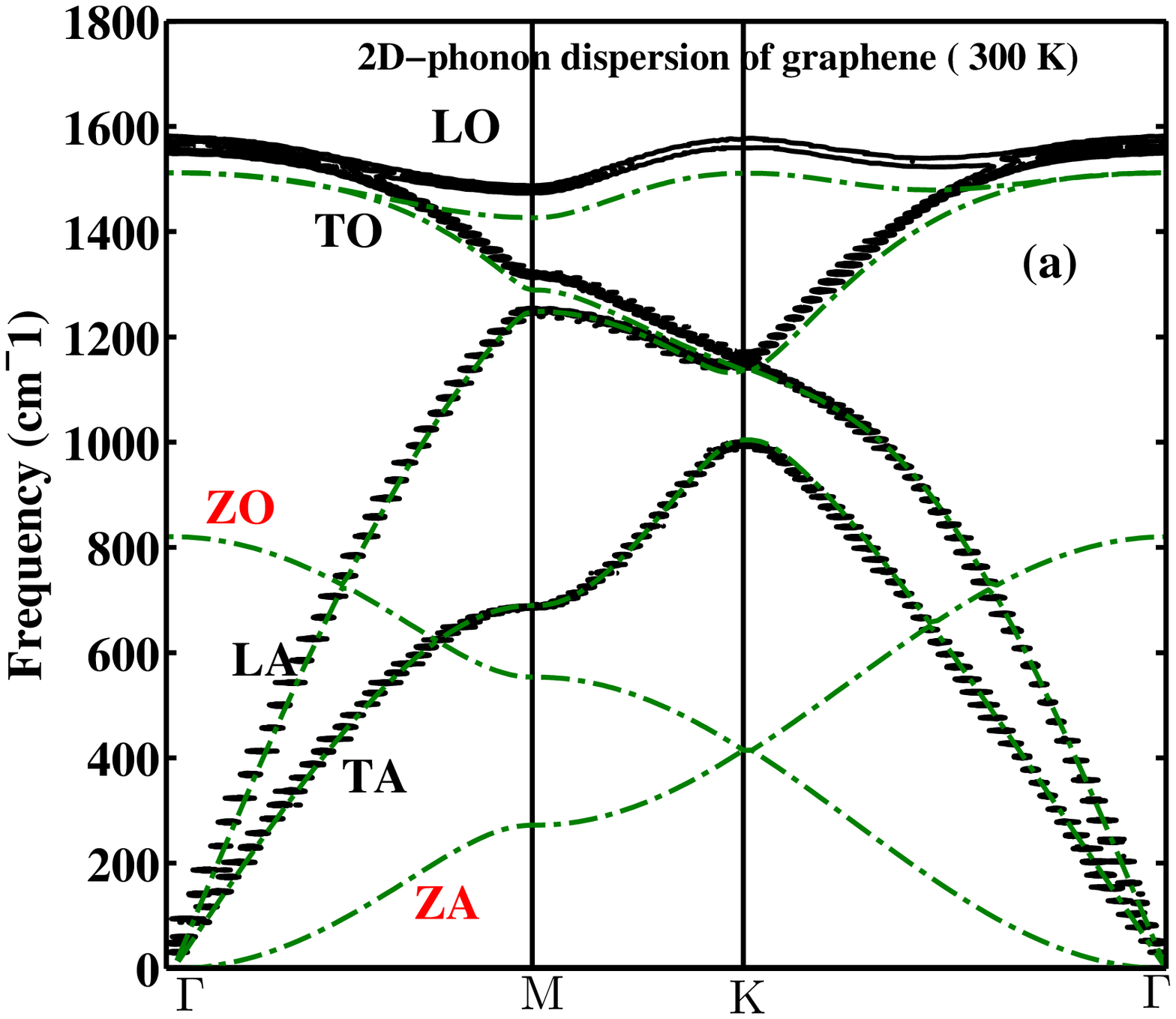}\includegraphics[scale=0.342]{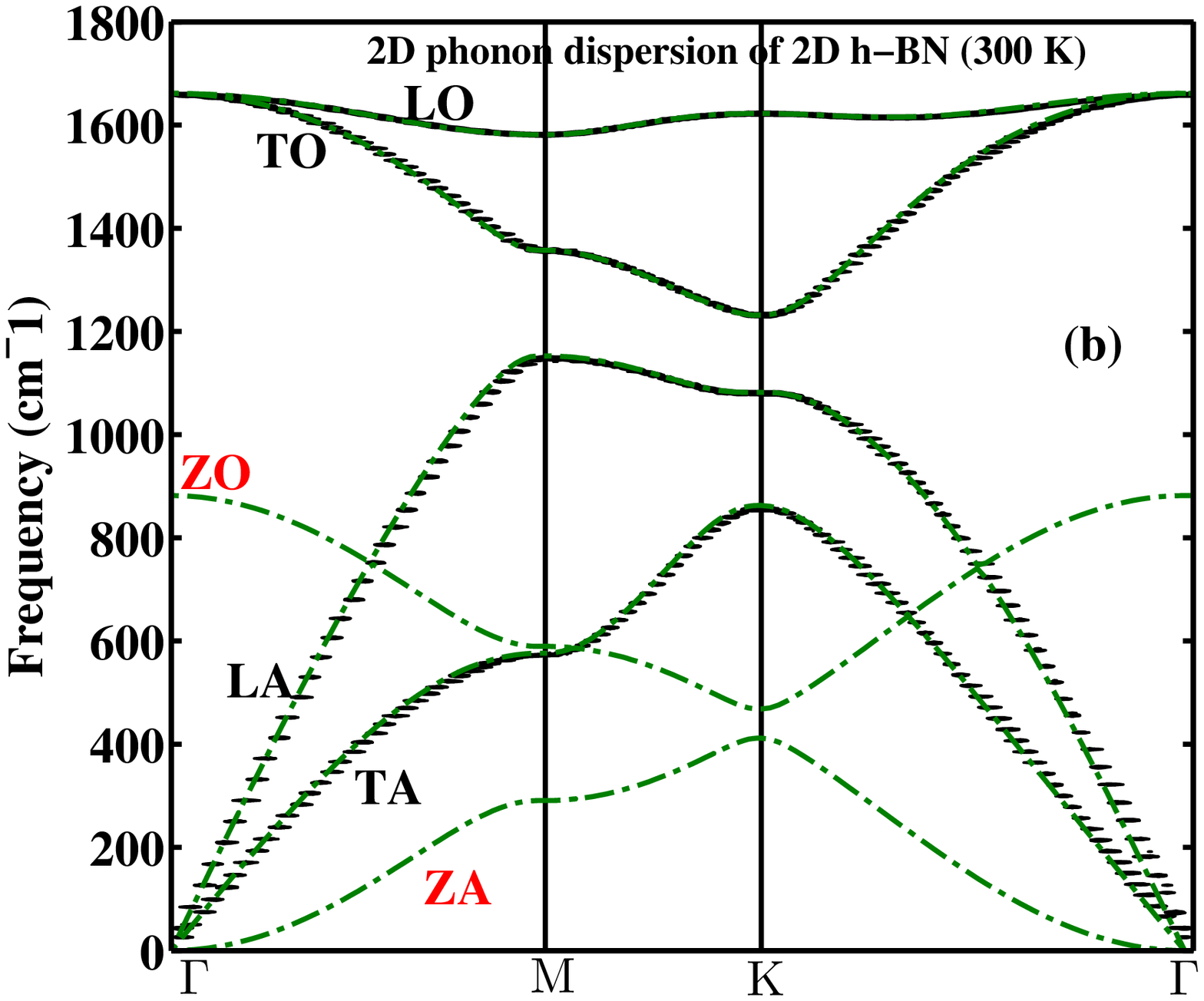}\includegraphics[scale=0.33]{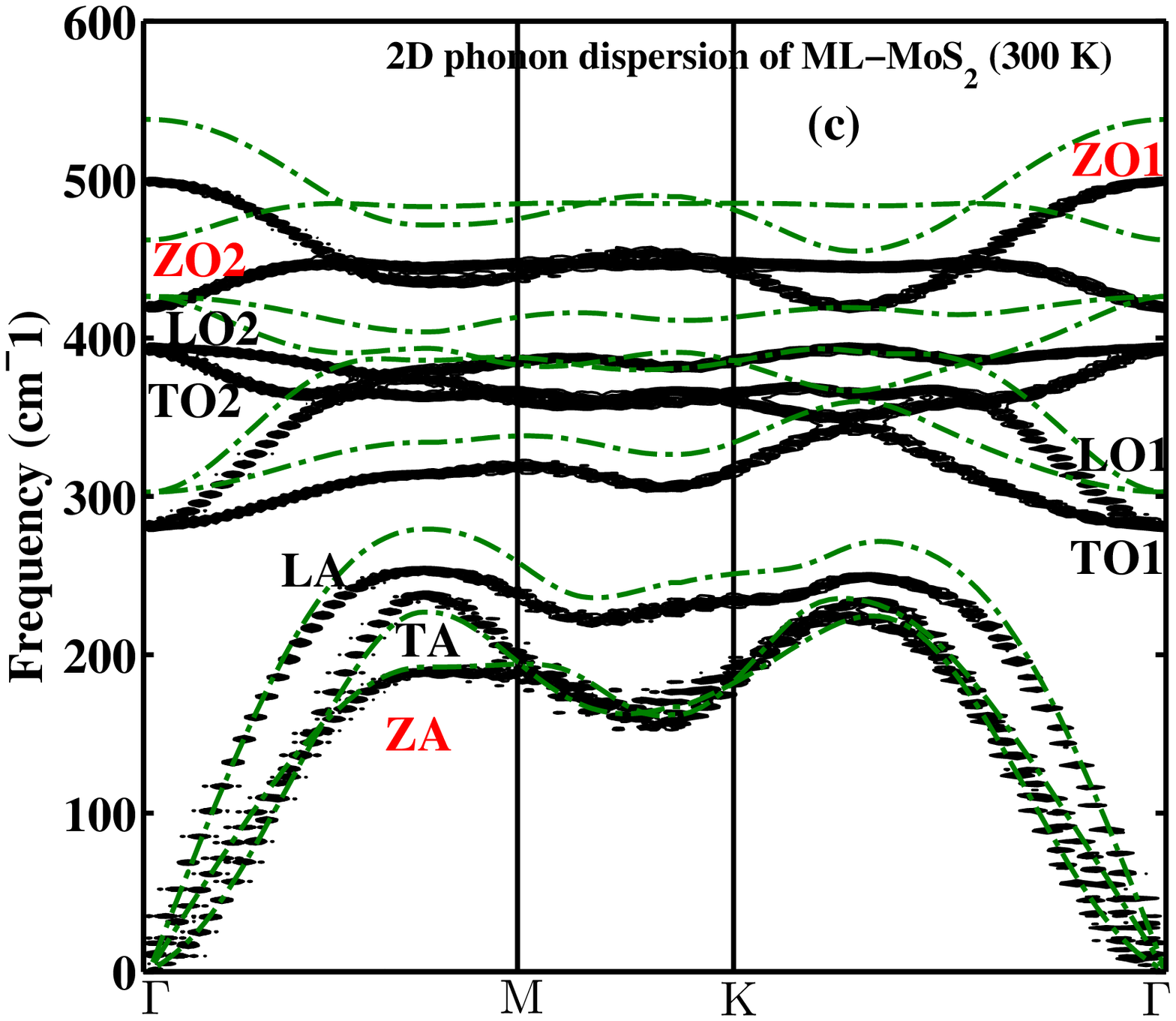}\protect\caption{\label{fig:The-dispersion-curves}The phonon dispersion of (a) graphene,
(b) 2D h-BN and (c) ML-MoS\protect\textsubscript{2}. The green-dot-dash
lines are obtained from lattice dynamics calculations (LD) at 0 K,
the thick black line is computed directly from 2D-molecular dynamics
simulations at 300 K. The out-of-plane modes (ZA, ZO) are absent in
2D phonon dispersion (obtained from MD simulations) of graphene and
2D-h-BN, as a results of constraining the out-of-plane motion. In
ML-MoS\protect\textsubscript{2 }, all modes are present in both LD
and 2D phonon dispersion curves, this attributes to its finite thickness
effects }
\end{figure*}

Figure \ref{fig:The-dispersion-curves} shows the 2D phonon dispersion
of above mentioned honeycomb structures. The green dot-dash curve
is obtained using LD method at 0K, the thick black curve is computed
directly from MD simulation. The 3D phonon dispersions of graphene
and 2D h-BN at finite temperatures are reported in our earlier papers\cite{2053-1583-2-3-035014,C5CP06111C},
hence here we are focusing on 2D phonon dispersion at 300 K. The graphene
and 2D h-BN unitcell contains 2 basis atom, which leads to six modes
of vibrations, three of them are acoustic (A) and remaining three
are optic (O) modes. The modes are labeled according to their polarizations,
the letter 'L', 'T', and 'Z' are used to denote longitudinal, transverse
and out-of-plane modes respectively. The ZA mode shows a quadratic
dispersion in graphene and 2D h-BN, which is a characteristic feature
of layered compounds\cite{Lifshitz1952}, and it is due to $D_{3h}$
point group symmetry\cite{R.Saito1998}. The Overall agreement of
LD frequencies of graphene and h-BN with previous calculations are
satisfactory \cite{Karssemeijer20111611,PhysRevB.84.085409}. In 2D
dispersion, the in-plane acoustic modes LA and TA shows similar behavior
as reported in 3D dispersion\cite{2053-1583-2-3-035014,C5CP06111C}.
The most interesting phenomena observed here is the absence of out-of-plane
modes such as ZA and ZO in 2D phonon dispersions of both graphene
and 2D h-BN. Since we arrested the motion of atoms along Z directions,
the branches corresponding to out-of-plane motions are missing in
phonon dispersion. This complete absence of ZA and ZO mode is the
reason behind the continuous thermal expansion of \textbf{\textit{a}}-lattice
in 2D simulations, and this observation is completely agreeing with
Grüneisen theory based analysis. The novelty of the present approach
is that, it exposes the importance of out-of-plane modes (ZA) in determining
the thermal expansion behavior, by computing the phonon dispersion
from the typical dynamics of atoms, instead of symmetry based LD methods. 

In ML-MoS\textsubscript{2}, due to the trigonal prismatic arrangement
of Mo and S atoms, vibrational modes behaves quite differently from
graphene and 2D h-BN. The unitcell of ML-MoS\textsubscript{2} contains
three basis atoms, hence there will be nine modes of vibrations (3
acoustic+6 optic). Figure \ref{fig:The-dispersion-curves}c displays
the phonon dispersion in h-MoS\textsubscript{2}. The LD calculations
are in good agreement with previous reports\cite{PhysRevB.89.035438,PhysRevB.84.155413}.
The gap between the acoustic and optic mode (TO1) is discernible,
where three acoustic branches TA, LA and ZA are separated below the
optic branch (TO1) by \textasciitilde{}55 cm\textsuperscript{\textminus 1}
at M point in the Brillouin zone, and it is in agreement with \textit{ab
initio}\textit{\textcolor{red}{{} }}calculations\cite{PhysRevB.89.035438}.
The LA and TA modes show linear dispersion, while ZA mode exhibits
quadratic dispersion around the $\Gamma$ point, analogous to graphene
and h-BN. Unlike graphene and h-BN, all out-of-plane modes are present
in 2D dispersion of ML-MoS\textsubscript{2}. Though we arrested the
out-of-plane motion, the ZA and ZO branches are still persisting in
ML-MoS\textsubscript{2}, and this can be ascribed to the finite thickness
effect of ML-MoS\textsubscript{2}. The graphene and h-BN are one
atom thick structure and have more flexibility along out-of-plane
direction, the S-Mo-S sandwich structure of ML-MoS\textsubscript{2}
makes it a more rigid material along out-of-plane direction and leads
to less rippling. The magnitude of thermally excited ripples can be
quantified using the height-height correlation function $\bigl\langle h^{2}\bigr\rangle$,
and its value is much smaller for ML-MoS\textsubscript{2 }in comparison
with graphene, and it is an outcome of less rippling behaviour of
ML-MoS\textsubscript{2}\cite{PhysRevB.91.014101}. The finite thickness
of ML-MoS\textsubscript{2} counteracts the membrane effects, and
hence the origin of bending mode (ZA) is not purely due to the out-of-plane
vibrations as in graphene and h-BN.

\section{Conclusions}

Thermally excited ripples are inevitable in 2D crystals, and they
can affect the thermo-physical properties of these materials significantly.
Inorder to delineate the role of ripples on thermal expansion of 2D
honeycomb materials (graphene, 2D h-BN and ML-MoS\textsubscript{2})
we performed three-dimensional (3D) and two dimensional (2D) molecular
dynamics simulations, the later cannot incorporate the effects of
ripples. The in-plane lattice parameter (\textbf{\textit{a}}-lattice)
of free-standing graphene calculated from 3D simulations shows a thermal
contraction upto T = 1300 K - 1400 K (depend on system size) and expands
thereafter. The linear thermal expansion coefficient (LTEC) changes
its sign from negative to positive in the above temperature range.
At the same time, the \textbf{\textit{a}}-lattice of very same system
obtained from 2D simulations shows continuous thermal expansion instead
of thermal contraction and LTECs are positive for all system sizes.
The above analysis was extended to 2D h-BN and found the similar discrepancy
between 2D and 3D simulations. Contradicting to graphene and 2D h-BN,
the \textbf{\textit{a}}-lattice of ML-MoS\textsubscript{2} shows
thermal expansion in both 3D and 2D simulations, and the LTECs are
positive and their system size dependence is marginal. The above discrepancy
is analyzed by computing the 2D phonon dispersion at 300 K using spectral
energy density method. The out-of-plane bending (ZA) mode is missing
in 2D phonon dispersions of graphene and 2D h-BN. The ZA mode, which
is responsible for thermal contraction of in-plane lattice parameter
at low temperature is absent in 2D simulations, which leads to continuous
thermal expansion. However, these modes are present in 2D dispersion
of ML-MoS\textsubscript{2}, indicates that its origin is not purely
due to the out-of-plane vibrations, and its effects on thermal expansion
is not significant as found in graphene and 2D h-BN systems. 

\bibliographystyle{apsrev}

\begin{thebibliography}{54}
\expandafter\ifx\csname natexlab\endcsname\relax\def\natexlab#1{#1}\fi
\expandafter\ifx\csname bibnamefont\endcsname\relax
  \def\bibnamefont#1{#1}\fi
\expandafter\ifx\csname bibfnamefont\endcsname\relax
  \def\bibfnamefont#1{#1}\fi
\expandafter\ifx\csname citenamefont\endcsname\relax
  \def\citenamefont#1{#1}\fi
\expandafter\ifx\csname url\endcsname\relax
  \def\url#1{\texttt{#1}}\fi
\expandafter\ifx\csname urlprefix\endcsname\relax\def\urlprefix{URL }\fi
\providecommand{\bibinfo}[2]{#2}
\providecommand{\eprint}[2][]{\url{#2}}

\bibitem[{\citenamefont{Geim and Novoselov}(2007)}]{geim2007rise}
\bibinfo{author}{\bibfnamefont{A.~K.} \bibnamefont{Geim}} \bibnamefont{and}
  \bibinfo{author}{\bibfnamefont{K.~S.} \bibnamefont{Novoselov}},
  \bibinfo{journal}{Nature materials} \textbf{\bibinfo{volume}{6}},
  \bibinfo{pages}{183} (\bibinfo{year}{2007}).

\bibitem[{\citenamefont{Balandin}(2011)}]{Balandin2011}
\bibinfo{author}{\bibfnamefont{A.~A.} \bibnamefont{Balandin}},
  \bibinfo{journal}{Nature materials} \textbf{\bibinfo{volume}{10}},
  \bibinfo{pages}{569} (\bibinfo{year}{2011}).

\bibitem[{\citenamefont{Lee et~al.}(2008)\citenamefont{Lee, Wei, Kysar, and
  Hone}}]{Lee18072008}
\bibinfo{author}{\bibfnamefont{C.}~\bibnamefont{Lee}},
  \bibinfo{author}{\bibfnamefont{X.}~\bibnamefont{Wei}},
  \bibinfo{author}{\bibfnamefont{J.~W.} \bibnamefont{Kysar}}, \bibnamefont{and}
  \bibinfo{author}{\bibfnamefont{J.}~\bibnamefont{Hone}},
  \bibinfo{journal}{Science} \textbf{\bibinfo{volume}{321}},
  \bibinfo{pages}{385} (\bibinfo{year}{2008}).

\bibitem[{\citenamefont{Schwierz}(2010)}]{Schwierz2010}
\bibinfo{author}{\bibfnamefont{F.}~\bibnamefont{Schwierz}},
  \bibinfo{journal}{Nat Nano} \textbf{\bibinfo{volume}{5}},
  \bibinfo{pages}{487} (\bibinfo{year}{2010}).

\bibitem[{\citenamefont{Castro~Neto et~al.}(2009)\citenamefont{Castro~Neto,
  Guinea, Peres, Novoselov, and Geim}}]{RevModPhys.81.109}
\bibinfo{author}{\bibfnamefont{A.~H.} \bibnamefont{Castro~Neto}},
  \bibinfo{author}{\bibfnamefont{F.}~\bibnamefont{Guinea}},
  \bibinfo{author}{\bibfnamefont{N.~M.~R.} \bibnamefont{Peres}},
  \bibinfo{author}{\bibfnamefont{K.~S.} \bibnamefont{Novoselov}},
  \bibnamefont{and} \bibinfo{author}{\bibfnamefont{A.~K.} \bibnamefont{Geim}},
  \bibinfo{journal}{Rev. Mod. Phys.} \textbf{\bibinfo{volume}{81}},
  \bibinfo{pages}{109} (\bibinfo{year}{2009}).

\bibitem[{\citenamefont{Golberg et~al.}(2010)\citenamefont{Golberg, Bando,
  Huang, Terao, Mitome, Tang, and Zhi}}]{doi:10.1021/nn1006495}
\bibinfo{author}{\bibfnamefont{D.}~\bibnamefont{Golberg}},
  \bibinfo{author}{\bibfnamefont{Y.}~\bibnamefont{Bando}},
  \bibinfo{author}{\bibfnamefont{Y.}~\bibnamefont{Huang}},
  \bibinfo{author}{\bibfnamefont{T.}~\bibnamefont{Terao}},
  \bibinfo{author}{\bibfnamefont{M.}~\bibnamefont{Mitome}},
  \bibinfo{author}{\bibfnamefont{C.}~\bibnamefont{Tang}}, \bibnamefont{and}
  \bibinfo{author}{\bibfnamefont{C.}~\bibnamefont{Zhi}}, \bibinfo{journal}{ACS
  Nano} \textbf{\bibinfo{volume}{4}}, \bibinfo{pages}{2979}
  (\bibinfo{year}{2010}).

\bibitem[{\citenamefont{Shi et~al.}(2010)\citenamefont{Shi, Hamsen, Jia, Kim,
  Reina, Hofmann, Hsu, Zhang, Li, Juang et~al.}}]{doi:10.1021/nl1023707}
\bibinfo{author}{\bibfnamefont{Y.}~\bibnamefont{Shi}},
  \bibinfo{author}{\bibfnamefont{C.}~\bibnamefont{Hamsen}},
  \bibinfo{author}{\bibfnamefont{X.}~\bibnamefont{Jia}},
  \bibinfo{author}{\bibfnamefont{K.~K.} \bibnamefont{Kim}},
  \bibinfo{author}{\bibfnamefont{A.}~\bibnamefont{Reina}},
  \bibinfo{author}{\bibfnamefont{M.}~\bibnamefont{Hofmann}},
  \bibinfo{author}{\bibfnamefont{A.~L.} \bibnamefont{Hsu}},
  \bibinfo{author}{\bibfnamefont{K.}~\bibnamefont{Zhang}},
  \bibinfo{author}{\bibfnamefont{H.}~\bibnamefont{Li}},
  \bibinfo{author}{\bibfnamefont{Z.-Y.} \bibnamefont{Juang}},
  \bibnamefont{et~al.}, \bibinfo{journal}{Nano Letters}
  \textbf{\bibinfo{volume}{10}}, \bibinfo{pages}{4134} (\bibinfo{year}{2010}).

\bibitem[{\citenamefont{Watanabe et~al.}(2004)\citenamefont{Watanabe,
  Taniguchi, and Kanda}}]{Watanabe2004}
\bibinfo{author}{\bibfnamefont{K.}~\bibnamefont{Watanabe}},
  \bibinfo{author}{\bibfnamefont{T.}~\bibnamefont{Taniguchi}},
  \bibnamefont{and} \bibinfo{author}{\bibfnamefont{H.}~\bibnamefont{Kanda}},
  \bibinfo{journal}{Nat Mater} \textbf{\bibinfo{volume}{3}},
  \bibinfo{pages}{404} (\bibinfo{year}{2004}).

\bibitem[{\citenamefont{Geim and Grigorieva}(2013)}]{Geim2013}
\bibinfo{author}{\bibfnamefont{A.~K.} \bibnamefont{Geim}} \bibnamefont{and}
  \bibinfo{author}{\bibfnamefont{I.~V.} \bibnamefont{Grigorieva}},
  \bibinfo{journal}{Nature} \textbf{\bibinfo{volume}{499}},
  \bibinfo{pages}{419} (\bibinfo{year}{2013}).

\bibitem[{\citenamefont{Chhowalla et~al.}(2013)\citenamefont{Chhowalla, Shin,
  Eda, Li, Loh, and Zhang}}]{Chhowalla2013}
\bibinfo{author}{\bibfnamefont{M.}~\bibnamefont{Chhowalla}},
  \bibinfo{author}{\bibfnamefont{H.~S.} \bibnamefont{Shin}},
  \bibinfo{author}{\bibfnamefont{G.}~\bibnamefont{Eda}},
  \bibinfo{author}{\bibfnamefont{L.-J.} \bibnamefont{Li}},
  \bibinfo{author}{\bibfnamefont{K.~P.} \bibnamefont{Loh}}, \bibnamefont{and}
  \bibinfo{author}{\bibfnamefont{H.}~\bibnamefont{Zhang}},
  \bibinfo{journal}{Nat Chem} \textbf{\bibinfo{volume}{5}},
  \bibinfo{pages}{263} (\bibinfo{year}{2013}).

\bibitem[{\citenamefont{Huang et~al.}(2013)\citenamefont{Huang, Zeng, and
  Zhang}}]{C2CS35387C}
\bibinfo{author}{\bibfnamefont{X.}~\bibnamefont{Huang}},
  \bibinfo{author}{\bibfnamefont{Z.}~\bibnamefont{Zeng}}, \bibnamefont{and}
  \bibinfo{author}{\bibfnamefont{H.}~\bibnamefont{Zhang}},
  \bibinfo{journal}{Chem. Soc. Rev.} \textbf{\bibinfo{volume}{42}},
  \bibinfo{pages}{1934} (\bibinfo{year}{2013}).

\bibitem[{\citenamefont{Splendiani et~al.}(2010)\citenamefont{Splendiani, Sun,
  Zhang, Li, Kim, Chim, Galli, and Wang}}]{doi:10.1021/nl903868w}
\bibinfo{author}{\bibfnamefont{A.}~\bibnamefont{Splendiani}},
  \bibinfo{author}{\bibfnamefont{L.}~\bibnamefont{Sun}},
  \bibinfo{author}{\bibfnamefont{Y.}~\bibnamefont{Zhang}},
  \bibinfo{author}{\bibfnamefont{T.}~\bibnamefont{Li}},
  \bibinfo{author}{\bibfnamefont{J.}~\bibnamefont{Kim}},
  \bibinfo{author}{\bibfnamefont{C.-Y.} \bibnamefont{Chim}},
  \bibinfo{author}{\bibfnamefont{G.}~\bibnamefont{Galli}}, \bibnamefont{and}
  \bibinfo{author}{\bibfnamefont{F.}~\bibnamefont{Wang}},
  \bibinfo{journal}{Nano Letters} \textbf{\bibinfo{volume}{10}},
  \bibinfo{pages}{1271} (\bibinfo{year}{2010}).

\bibitem[{\citenamefont{RadisavljevicB
  et~al.}(2011)\citenamefont{RadisavljevicB, RadenovicA, BrivioJ, GiacomettiV,
  and KisA}}]{RadisavljevicB2011}
\bibinfo{author}{\bibnamefont{RadisavljevicB}},
  \bibinfo{author}{\bibnamefont{RadenovicA}},
  \bibinfo{author}{\bibnamefont{BrivioJ}},
  \bibinfo{author}{\bibnamefont{GiacomettiV}}, \bibnamefont{and}
  \bibinfo{author}{\bibnamefont{KisA}}, \bibinfo{journal}{Nat Nano}
  \textbf{\bibinfo{volume}{6}}, \bibinfo{pages}{147} (\bibinfo{year}{2011}).

\bibitem[{\citenamefont{Min et~al.}(2013)\citenamefont{Min, Lee, Choi, Park,
  Nam, Kim, Ryu, and Im}}]{C2NR33443G}
\bibinfo{author}{\bibfnamefont{S.-W.} \bibnamefont{Min}},
  \bibinfo{author}{\bibfnamefont{H.~S.} \bibnamefont{Lee}},
  \bibinfo{author}{\bibfnamefont{H.~J.} \bibnamefont{Choi}},
  \bibinfo{author}{\bibfnamefont{M.~K.} \bibnamefont{Park}},
  \bibinfo{author}{\bibfnamefont{T.}~\bibnamefont{Nam}},
  \bibinfo{author}{\bibfnamefont{H.}~\bibnamefont{Kim}},
  \bibinfo{author}{\bibfnamefont{S.}~\bibnamefont{Ryu}}, \bibnamefont{and}
  \bibinfo{author}{\bibfnamefont{S.}~\bibnamefont{Im}},
  \bibinfo{journal}{Nanoscale} \textbf{\bibinfo{volume}{5}},
  \bibinfo{pages}{548} (\bibinfo{year}{2013}).

\bibitem[{\citenamefont{Na et~al.}(2014)\citenamefont{Na, Joo, Shin, Huh, Kim,
  Piao, Jin, Jang, Choi, Shim et~al.}}]{C3NR04218A}
\bibinfo{author}{\bibfnamefont{J.}~\bibnamefont{Na}},
  \bibinfo{author}{\bibfnamefont{M.-K.} \bibnamefont{Joo}},
  \bibinfo{author}{\bibfnamefont{M.}~\bibnamefont{Shin}},
  \bibinfo{author}{\bibfnamefont{J.}~\bibnamefont{Huh}},
  \bibinfo{author}{\bibfnamefont{J.-S.} \bibnamefont{Kim}},
  \bibinfo{author}{\bibfnamefont{M.}~\bibnamefont{Piao}},
  \bibinfo{author}{\bibfnamefont{J.-E.} \bibnamefont{Jin}},
  \bibinfo{author}{\bibfnamefont{H.-K.} \bibnamefont{Jang}},
  \bibinfo{author}{\bibfnamefont{H.~J.} \bibnamefont{Choi}},
  \bibinfo{author}{\bibfnamefont{J.~H.} \bibnamefont{Shim}},
  \bibnamefont{et~al.}, \bibinfo{journal}{Nanoscale}
  \textbf{\bibinfo{volume}{6}}, \bibinfo{pages}{433} (\bibinfo{year}{2014}).

\bibitem[{\citenamefont{Mermin}(1968)}]{PhysRev.176.250}
\bibinfo{author}{\bibfnamefont{N.~D.} \bibnamefont{Mermin}},
  \bibinfo{journal}{Phys. Rev.} \textbf{\bibinfo{volume}{176}},
  \bibinfo{pages}{250} (\bibinfo{year}{1968}).

\bibitem[{\citenamefont{Fasolino et~al.}(2007)\citenamefont{Fasolino, Los, and
  Katsnelson}}]{Fasolino2007}
\bibinfo{author}{\bibfnamefont{A.}~\bibnamefont{Fasolino}},
  \bibinfo{author}{\bibfnamefont{J.}~\bibnamefont{Los}}, \bibnamefont{and}
  \bibinfo{author}{\bibfnamefont{M.~I.} \bibnamefont{Katsnelson}},
  \bibinfo{journal}{Nature materials} \textbf{\bibinfo{volume}{6}},
  \bibinfo{pages}{858} (\bibinfo{year}{2007}).

\bibitem[{\citenamefont{Anees et~al.}(2015)\citenamefont{Anees, Valsakumar, and
  Panigrahi}}]{2053-1583-2-3-035014}
\bibinfo{author}{\bibfnamefont{P.}~\bibnamefont{Anees}},
  \bibinfo{author}{\bibfnamefont{M.~C.} \bibnamefont{Valsakumar}},
  \bibnamefont{and} \bibinfo{author}{\bibfnamefont{B.~K.}
  \bibnamefont{Panigrahi}}, \bibinfo{journal}{2D Materials}
  \textbf{\bibinfo{volume}{2}}, \bibinfo{pages}{035014} (\bibinfo{year}{2015}).

\bibitem[{\citenamefont{Anees et~al.}(2016{\natexlab{a}})\citenamefont{Anees,
  Valsakumar, and Panigrahi}}]{C5CP06111C}
\bibinfo{author}{\bibfnamefont{P.}~\bibnamefont{Anees}},
  \bibinfo{author}{\bibfnamefont{M.~C.} \bibnamefont{Valsakumar}},
  \bibnamefont{and} \bibinfo{author}{\bibfnamefont{B.~K.}
  \bibnamefont{Panigrahi}}, \bibinfo{journal}{Phys. Chem. Chem. Phys.}
  \textbf{\bibinfo{volume}{18}}, \bibinfo{pages}{2672}
  (\bibinfo{year}{2016}{\natexlab{a}}).

\bibitem[{\citenamefont{Meyer et~al.}(2007)\citenamefont{Meyer, Geim,
  Katsnelson, Novoselov, Booth, and Roth}}]{Meyer2007}
\bibinfo{author}{\bibfnamefont{J.~C.} \bibnamefont{Meyer}},
  \bibinfo{author}{\bibfnamefont{A.~K.} \bibnamefont{Geim}},
  \bibinfo{author}{\bibfnamefont{M.~I.} \bibnamefont{Katsnelson}},
  \bibinfo{author}{\bibfnamefont{K.~S.} \bibnamefont{Novoselov}},
  \bibinfo{author}{\bibfnamefont{T.~J.} \bibnamefont{Booth}}, \bibnamefont{and}
  \bibinfo{author}{\bibfnamefont{S.}~\bibnamefont{Roth}},
  \bibinfo{journal}{Nature} \textbf{\bibinfo{volume}{446}}, \bibinfo{pages}{60}
  (\bibinfo{year}{2007}).

\bibitem[{\citenamefont{{Gallagher} et~al.}(2015)\citenamefont{{Gallagher},
  {Lee}, {Amet}, {Maksymovych}, {Wang}, {Wang}, {Lu}, {Zhang}, {Watanabe},
  {Taniguchi} et~al.}}]{2015arXiv150405253G}
\bibinfo{author}{\bibfnamefont{P.}~\bibnamefont{{Gallagher}}},
  \bibinfo{author}{\bibfnamefont{M.}~\bibnamefont{{Lee}}},
  \bibinfo{author}{\bibfnamefont{F.}~\bibnamefont{{Amet}}},
  \bibinfo{author}{\bibfnamefont{P.}~\bibnamefont{{Maksymovych}}},
  \bibinfo{author}{\bibfnamefont{J.}~\bibnamefont{{Wang}}},
  \bibinfo{author}{\bibfnamefont{S.}~\bibnamefont{{Wang}}},
  \bibinfo{author}{\bibfnamefont{X.}~\bibnamefont{{Lu}}},
  \bibinfo{author}{\bibfnamefont{G.}~\bibnamefont{{Zhang}}},
  \bibinfo{author}{\bibfnamefont{K.}~\bibnamefont{{Watanabe}}},
  \bibinfo{author}{\bibfnamefont{T.}~\bibnamefont{{Taniguchi}}},
  \bibnamefont{et~al.}, \bibinfo{journal}{ArXiv e-prints}
  (\bibinfo{year}{2015}).

\bibitem[{\citenamefont{{Pereira} and {Castro
  Neto}}(2008)}]{2008arXiv0810.4539P}
\bibinfo{author}{\bibfnamefont{V.~M.} \bibnamefont{{Pereira}}}
  \bibnamefont{and} \bibinfo{author}{\bibfnamefont{A.~H.} \bibnamefont{{Castro
  Neto}}}, \bibinfo{journal}{ArXiv e-prints}  (\bibinfo{year}{2008}).

\bibitem[{\citenamefont{Elias et~al.}(2009)\citenamefont{Elias, Nair,
  Mohiuddin, Morozov, Blake, Halsall, Ferrari, Boukhvalov, Katsnelson, Geim
  et~al.}}]{Elias610}
\bibinfo{author}{\bibfnamefont{D.~C.} \bibnamefont{Elias}},
  \bibinfo{author}{\bibfnamefont{R.~R.} \bibnamefont{Nair}},
  \bibinfo{author}{\bibfnamefont{T.~M.~G.} \bibnamefont{Mohiuddin}},
  \bibinfo{author}{\bibfnamefont{S.~V.} \bibnamefont{Morozov}},
  \bibinfo{author}{\bibfnamefont{P.}~\bibnamefont{Blake}},
  \bibinfo{author}{\bibfnamefont{M.~P.} \bibnamefont{Halsall}},
  \bibinfo{author}{\bibfnamefont{A.~C.} \bibnamefont{Ferrari}},
  \bibinfo{author}{\bibfnamefont{D.~W.} \bibnamefont{Boukhvalov}},
  \bibinfo{author}{\bibfnamefont{M.~I.} \bibnamefont{Katsnelson}},
  \bibinfo{author}{\bibfnamefont{A.~K.} \bibnamefont{Geim}},
  \bibnamefont{et~al.}, \bibinfo{journal}{Science}
  \textbf{\bibinfo{volume}{323}}, \bibinfo{pages}{610} (\bibinfo{year}{2009}).

\bibitem[{\citenamefont{Mounet and Marzari}(2005)}]{PhysRevB.71.205214}
\bibinfo{author}{\bibfnamefont{N.}~\bibnamefont{Mounet}} \bibnamefont{and}
  \bibinfo{author}{\bibfnamefont{N.}~\bibnamefont{Marzari}},
  \bibinfo{journal}{Phys. Rev. B} \textbf{\bibinfo{volume}{71}},
  \bibinfo{pages}{205214} (\bibinfo{year}{2005}).

\bibitem[{\citenamefont{Zakharchenko et~al.}(2009)\citenamefont{Zakharchenko,
  Katsnelson, and Fasolino}}]{PhysRevLett.102.046808}
\bibinfo{author}{\bibfnamefont{K.~V.} \bibnamefont{Zakharchenko}},
  \bibinfo{author}{\bibfnamefont{M.~I.} \bibnamefont{Katsnelson}},
  \bibnamefont{and} \bibinfo{author}{\bibfnamefont{A.}~\bibnamefont{Fasolino}},
  \bibinfo{journal}{Phys. Rev. Lett.} \textbf{\bibinfo{volume}{102}},
  \bibinfo{pages}{046808} (\bibinfo{year}{2009}).

\bibitem[{\citenamefont{Pozzo et~al.}(2011)\citenamefont{Pozzo, Alf\`e,
  Lacovig, Hofmann, Lizzit, and Baraldi}}]{PhysRevLett.106.135501}
\bibinfo{author}{\bibfnamefont{M.}~\bibnamefont{Pozzo}},
  \bibinfo{author}{\bibfnamefont{D.}~\bibnamefont{Alf\`e}},
  \bibinfo{author}{\bibfnamefont{P.}~\bibnamefont{Lacovig}},
  \bibinfo{author}{\bibfnamefont{P.}~\bibnamefont{Hofmann}},
  \bibinfo{author}{\bibfnamefont{S.}~\bibnamefont{Lizzit}}, \bibnamefont{and}
  \bibinfo{author}{\bibfnamefont{A.}~\bibnamefont{Baraldi}},
  \bibinfo{journal}{Phys. Rev. Lett.} \textbf{\bibinfo{volume}{106}},
  \bibinfo{pages}{135501} (\bibinfo{year}{2011}).

\bibitem[{\citenamefont{Sevik}(2014)}]{PhysRevB.89.035422}
\bibinfo{author}{\bibfnamefont{C.}~\bibnamefont{Sevik}},
  \bibinfo{journal}{Phys. Rev. B} \textbf{\bibinfo{volume}{89}},
  \bibinfo{pages}{035422} (\bibinfo{year}{2014}).

\bibitem[{\citenamefont{Bao et~al.}(2009)\citenamefont{Bao, Miao, Chen, Zhang,
  Jang, Dames, and Lau}}]{Bao2009}
\bibinfo{author}{\bibfnamefont{W.}~\bibnamefont{Bao}},
  \bibinfo{author}{\bibfnamefont{F.}~\bibnamefont{Miao}},
  \bibinfo{author}{\bibfnamefont{Z.}~\bibnamefont{Chen}},
  \bibinfo{author}{\bibfnamefont{H.}~\bibnamefont{Zhang}},
  \bibinfo{author}{\bibfnamefont{W.}~\bibnamefont{Jang}},
  \bibinfo{author}{\bibfnamefont{C.}~\bibnamefont{Dames}}, \bibnamefont{and}
  \bibinfo{author}{\bibfnamefont{C.~N.} \bibnamefont{Lau}},
  \bibinfo{journal}{Nat Nano} \textbf{\bibinfo{volume}{4}},
  \bibinfo{pages}{562} (\bibinfo{year}{2009}).

\bibitem[{\citenamefont{Singh et~al.}(2010)\citenamefont{Singh, Sengupta,
  Solanki, Dhall, Allain, Dhara, Pant, and Deshmukh}}]{0957-4484-21-16-165204}
\bibinfo{author}{\bibfnamefont{V.}~\bibnamefont{Singh}},
  \bibinfo{author}{\bibfnamefont{S.}~\bibnamefont{Sengupta}},
  \bibinfo{author}{\bibfnamefont{H.~S.} \bibnamefont{Solanki}},
  \bibinfo{author}{\bibfnamefont{R.}~\bibnamefont{Dhall}},
  \bibinfo{author}{\bibfnamefont{A.}~\bibnamefont{Allain}},
  \bibinfo{author}{\bibfnamefont{S.}~\bibnamefont{Dhara}},
  \bibinfo{author}{\bibfnamefont{P.}~\bibnamefont{Pant}}, \bibnamefont{and}
  \bibinfo{author}{\bibfnamefont{M.~M.} \bibnamefont{Deshmukh}},
  \bibinfo{journal}{Nanotechnology} \textbf{\bibinfo{volume}{21}},
  \bibinfo{pages}{165204} (\bibinfo{year}{2010}).

\bibitem[{\citenamefont{Yoon et~al.}(2011)\citenamefont{Yoon, Son, and
  Cheong}}]{doi:10.1021/nl201488g}
\bibinfo{author}{\bibfnamefont{D.}~\bibnamefont{Yoon}},
  \bibinfo{author}{\bibfnamefont{Y.-W.} \bibnamefont{Son}}, \bibnamefont{and}
  \bibinfo{author}{\bibfnamefont{H.}~\bibnamefont{Cheong}},
  \bibinfo{journal}{Nano Letters} \textbf{\bibinfo{volume}{11}},
  \bibinfo{pages}{3227} (\bibinfo{year}{2011}).

\bibitem[{\citenamefont{Pan et~al.}(2012)\citenamefont{Pan, Xiao, Zhu, Yu,
  Zhang, Ni, Watanabe, Taniguchi, Shi, and Wang}}]{Pan2012}
\bibinfo{author}{\bibfnamefont{W.}~\bibnamefont{Pan}},
  \bibinfo{author}{\bibfnamefont{J.}~\bibnamefont{Xiao}},
  \bibinfo{author}{\bibfnamefont{J.}~\bibnamefont{Zhu}},
  \bibinfo{author}{\bibfnamefont{C.}~\bibnamefont{Yu}},
  \bibinfo{author}{\bibfnamefont{G.}~\bibnamefont{Zhang}},
  \bibinfo{author}{\bibfnamefont{Z.}~\bibnamefont{Ni}},
  \bibinfo{author}{\bibfnamefont{K.}~\bibnamefont{Watanabe}},
  \bibinfo{author}{\bibfnamefont{T.}~\bibnamefont{Taniguchi}},
  \bibinfo{author}{\bibfnamefont{Y.}~\bibnamefont{Shi}}, \bibnamefont{and}
  \bibinfo{author}{\bibfnamefont{X.}~\bibnamefont{Wang}},
  \bibinfo{journal}{Scientific Reports} \textbf{\bibinfo{volume}{2}},
  \bibinfo{pages}{893 EP } (\bibinfo{year}{2012}).

\bibitem[{\citenamefont{Plimpton}(1995)}]{Plimpton19951}
\bibinfo{author}{\bibfnamefont{S.}~\bibnamefont{Plimpton}},
  \bibinfo{journal}{Journal of Computational Physics}
  \textbf{\bibinfo{volume}{117}}, \bibinfo{pages}{1 } (\bibinfo{year}{1995}),
  \urlprefix\url{http://lammps.sandia.gov}.

\bibitem[{\citenamefont{Los and Fasolino}(2003)}]{PhysRevB.68.024107}
\bibinfo{author}{\bibfnamefont{J.~H.} \bibnamefont{Los}} \bibnamefont{and}
  \bibinfo{author}{\bibfnamefont{A.}~\bibnamefont{Fasolino}},
  \bibinfo{journal}{Phys. Rev. B} \textbf{\bibinfo{volume}{68}},
  \bibinfo{pages}{024107} (\bibinfo{year}{2003}).

\bibitem[{\citenamefont{Jiang et~al.}(2009)\citenamefont{Jiang, Wang, and
  Li}}]{PhysRevB.80.205429}
\bibinfo{author}{\bibfnamefont{J.-W.} \bibnamefont{Jiang}},
  \bibinfo{author}{\bibfnamefont{J.-S.} \bibnamefont{Wang}}, \bibnamefont{and}
  \bibinfo{author}{\bibfnamefont{B.}~\bibnamefont{Li}}, \bibinfo{journal}{Phys.
  Rev. B} \textbf{\bibinfo{volume}{80}}, \bibinfo{pages}{205429}
  (\bibinfo{year}{2009}).

\bibitem[{\citenamefont{Sevik et~al.}(2011)\citenamefont{Sevik, Kinaci,
  Haskins, and \ifmmode \mbox{\c{C}}\else \c{C}\fi{}a\ifmmode \breve{g}\else
  \u{g}\fi{}\ifmmode \imath \else~\i \fi{}n}}]{PhysRevB.84.085409}
\bibinfo{author}{\bibfnamefont{C.}~\bibnamefont{Sevik}},
  \bibinfo{author}{\bibfnamefont{A.}~\bibnamefont{Kinaci}},
  \bibinfo{author}{\bibfnamefont{J.~B.} \bibnamefont{Haskins}},
  \bibnamefont{and} \bibinfo{author}{\bibfnamefont{T.}~\bibnamefont{\ifmmode
  \mbox{\c{C}}\else \c{C}\fi{}a\ifmmode \breve{g}\else \u{g}\fi{}\ifmmode
  \imath \else~\i \fi{}n}}, \bibinfo{journal}{Phys. Rev. B}
  \textbf{\bibinfo{volume}{84}}, \bibinfo{pages}{085409}
  (\bibinfo{year}{2011}).

\bibitem[{\citenamefont{Pacilé et~al.}(2008)\citenamefont{Pacilé, Meyer, Girit,
  and Zettl}}]{Pacile2008}
\bibinfo{author}{\bibfnamefont{D.}~\bibnamefont{Pacilé}},
  \bibinfo{author}{\bibfnamefont{J.~C.} \bibnamefont{Meyer}},
  \bibinfo{author}{\bibfnamefont{Ç.~Ö.} \bibnamefont{Girit}}, \bibnamefont{and}
  \bibinfo{author}{\bibfnamefont{A.}~\bibnamefont{Zettl}},
  \bibinfo{journal}{Applied Physics Letters} \textbf{\bibinfo{volume}{92}},
  \bibinfo{pages}{133107} (\bibinfo{year}{2008}).

\bibitem[{\citenamefont{Nag et~al.}(2010)\citenamefont{Nag, Raidongia, Hembram,
  Datta, Waghmare, and Rao}}]{doi:10.1021/nn9018762}
\bibinfo{author}{\bibfnamefont{A.}~\bibnamefont{Nag}},
  \bibinfo{author}{\bibfnamefont{K.}~\bibnamefont{Raidongia}},
  \bibinfo{author}{\bibfnamefont{K.~P. S.~S.} \bibnamefont{Hembram}},
  \bibinfo{author}{\bibfnamefont{R.}~\bibnamefont{Datta}},
  \bibinfo{author}{\bibfnamefont{U.~V.} \bibnamefont{Waghmare}},
  \bibnamefont{and} \bibinfo{author}{\bibfnamefont{C.~N.~R.}
  \bibnamefont{Rao}}, \bibinfo{journal}{ACS Nano} \textbf{\bibinfo{volume}{4}},
  \bibinfo{pages}{1539} (\bibinfo{year}{2010}).

\bibitem[{\citenamefont{Paszkowicz et~al.}(2002)\citenamefont{Paszkowicz,
  Pelka, Knapp, Szyszko, and Podsiadlo}}]{raey}
\bibinfo{author}{\bibfnamefont{W.}~\bibnamefont{Paszkowicz}},
  \bibinfo{author}{\bibfnamefont{J.}~\bibnamefont{Pelka}},
  \bibinfo{author}{\bibfnamefont{M.}~\bibnamefont{Knapp}},
  \bibinfo{author}{\bibfnamefont{T.}~\bibnamefont{Szyszko}}, \bibnamefont{and}
  \bibinfo{author}{\bibfnamefont{S.}~\bibnamefont{Podsiadlo}},
  \bibinfo{journal}{Applied Physics A} \textbf{\bibinfo{volume}{75}},
  \bibinfo{pages}{431} (\bibinfo{year}{2002}).

\bibitem[{\citenamefont{Liang et~al.}(2009)\citenamefont{Liang, Phillpot, and
  Sinnott}}]{PhysRevB.79.245110}
\bibinfo{author}{\bibfnamefont{T.}~\bibnamefont{Liang}},
  \bibinfo{author}{\bibfnamefont{S.~R.} \bibnamefont{Phillpot}},
  \bibnamefont{and} \bibinfo{author}{\bibfnamefont{S.~B.}
  \bibnamefont{Sinnott}}, \bibinfo{journal}{Phys. Rev. B}
  \textbf{\bibinfo{volume}{79}}, \bibinfo{pages}{245110}
  (\bibinfo{year}{2009}).

\bibitem[{\citenamefont{Stewart and Spearot}(2013)}]{0965-0393-21-4-045003}
\bibinfo{author}{\bibfnamefont{J.~A.} \bibnamefont{Stewart}} \bibnamefont{and}
  \bibinfo{author}{\bibfnamefont{D.~E.} \bibnamefont{Spearot}},
  \bibinfo{journal}{Modelling and Simulation in Materials Science and
  Engineering} \textbf{\bibinfo{volume}{21}}, \bibinfo{pages}{045003}
  (\bibinfo{year}{2013}).

\bibitem[{\citenamefont{Ataca et~al.}(2011)\citenamefont{Ataca, Topsakal,
  Aktürk, and Ciraci}}]{doi:10.1021/jp205116x}
\bibinfo{author}{\bibfnamefont{C.}~\bibnamefont{Ataca}},
  \bibinfo{author}{\bibfnamefont{M.}~\bibnamefont{Topsakal}},
  \bibinfo{author}{\bibfnamefont{E.}~\bibnamefont{Aktürk}}, \bibnamefont{and}
  \bibinfo{author}{\bibfnamefont{S.}~\bibnamefont{Ciraci}},
  \bibinfo{journal}{The Journal of Physical Chemistry C}
  \textbf{\bibinfo{volume}{115}}, \bibinfo{pages}{16354}
  (\bibinfo{year}{2011}).

\bibitem[{\citenamefont{Murray and Evans}(1979)}]{JCR:JCRA18163}
\bibinfo{author}{\bibfnamefont{R.}~\bibnamefont{Murray}} \bibnamefont{and}
  \bibinfo{author}{\bibfnamefont{B.}~\bibnamefont{Evans}},
  \bibinfo{journal}{Journal of Applied Crystallography}
  \textbf{\bibinfo{volume}{12}}, \bibinfo{pages}{312} (\bibinfo{year}{1979}).

\bibitem[{\citenamefont{Huang et~al.}(2014)\citenamefont{Huang, Gong, and
  Zeng}}]{PhysRevB.90.045409}
\bibinfo{author}{\bibfnamefont{L.~F.} \bibnamefont{Huang}},
  \bibinfo{author}{\bibfnamefont{P.~L.} \bibnamefont{Gong}}, \bibnamefont{and}
  \bibinfo{author}{\bibfnamefont{Z.}~\bibnamefont{Zeng}},
  \bibinfo{journal}{Phys. Rev. B} \textbf{\bibinfo{volume}{90}},
  \bibinfo{pages}{045409} (\bibinfo{year}{2014}).

\bibitem[{\citenamefont{Anees et~al.}(2016{\natexlab{b}})\citenamefont{Anees,
  Valsakumar, and
  Panigrahi}}]{:/content/aip/journal/apl/108/10/10.1063/1.4943546}
\bibinfo{author}{\bibfnamefont{P.}~\bibnamefont{Anees}},
  \bibinfo{author}{\bibfnamefont{M.~C.} \bibnamefont{Valsakumar}},
  \bibnamefont{and} \bibinfo{author}{\bibfnamefont{B.~K.}
  \bibnamefont{Panigrahi}}, \bibinfo{journal}{Applied Physics Letters}
  \textbf{\bibinfo{volume}{108}} (\bibinfo{year}{2016}{\natexlab{b}}).

\bibitem[{\citenamefont{El-Mahalawy and Evans}(1976)}]{JCR:JCRA13982}
\bibinfo{author}{\bibfnamefont{S.~H.} \bibnamefont{El-Mahalawy}}
  \bibnamefont{and} \bibinfo{author}{\bibfnamefont{B.~L.} \bibnamefont{Evans}},
  \bibinfo{journal}{Journal of Applied Crystallography}
  \textbf{\bibinfo{volume}{9}}, \bibinfo{pages}{403} (\bibinfo{year}{1976}).

\bibitem[{\citenamefont{Singh et~al.}(2015)\citenamefont{Singh, Neek-Amal,
  Costamagna, and Peeters}}]{PhysRevB.91.014101}
\bibinfo{author}{\bibfnamefont{S.~K.} \bibnamefont{Singh}},
  \bibinfo{author}{\bibfnamefont{M.}~\bibnamefont{Neek-Amal}},
  \bibinfo{author}{\bibfnamefont{S.}~\bibnamefont{Costamagna}},
  \bibnamefont{and} \bibinfo{author}{\bibfnamefont{F.~M.}
  \bibnamefont{Peeters}}, \bibinfo{journal}{Phys. Rev. B}
  \textbf{\bibinfo{volume}{91}}, \bibinfo{pages}{014101}
  (\bibinfo{year}{2015}).

\bibitem[{\citenamefont{Schelling and Keblinski}(2003)}]{PhysRevB.68.035425}
\bibinfo{author}{\bibfnamefont{P.~K.} \bibnamefont{Schelling}}
  \bibnamefont{and}
  \bibinfo{author}{\bibfnamefont{P.}~\bibnamefont{Keblinski}},
  \bibinfo{journal}{Phys. Rev. B} \textbf{\bibinfo{volume}{68}},
  \bibinfo{pages}{035425} (\bibinfo{year}{2003}).

\bibitem[{\citenamefont{Lifshitz}(1952)}]{Lifshitz1952}
\bibinfo{author}{\bibfnamefont{M.}~\bibnamefont{Lifshitz}},
  \bibinfo{journal}{Zh. Eksp. Teor. Fiz. 22, 475}  (\bibinfo{year}{1952}).

\bibitem[{\citenamefont{Karssemeijer and Fasolino}(2010)}]{Karssemeijer2010}
\bibinfo{author}{\bibfnamefont{L.}~\bibnamefont{Karssemeijer}}
  \bibnamefont{and} \bibinfo{author}{\bibfnamefont{A.}~\bibnamefont{Fasolino}},
  Master's thesis, \bibinfo{school}{Theory of Condensed Matter, Institute for
  Molecules and Materials,Radboud University Nijmegen} (\bibinfo{year}{2010}).

\bibitem[{\citenamefont{Dove}(1993)}]{Dove1993}
\bibinfo{author}{\bibfnamefont{M.~T.} \bibnamefont{Dove}},
  \emph{\bibinfo{title}{Introduction to Lattice Dynamics}}
  (\bibinfo{publisher}{Cambridge University Press}, \bibinfo{year}{1993}).

\bibitem[{\citenamefont{R.~Saito and Dresselhaus}(1998)}]{R.Saito1998}
\bibinfo{author}{\bibfnamefont{G.~D.} \bibnamefont{R.~Saito}} \bibnamefont{and}
  \bibinfo{author}{\bibfnamefont{M.}~\bibnamefont{Dresselhaus}},
  \emph{\bibinfo{title}{Physical Properties of Carbon Nanotubes}}
  (\bibinfo{publisher}{London: Imperial College Press}, \bibinfo{year}{1998}).

\bibitem[{\citenamefont{Karssemeijer and
  Fasolino}(2011)}]{Karssemeijer20111611}
\bibinfo{author}{\bibfnamefont{L.}~\bibnamefont{Karssemeijer}}
  \bibnamefont{and} \bibinfo{author}{\bibfnamefont{A.}~\bibnamefont{Fasolino}},
  \bibinfo{journal}{Surface Science} \textbf{\bibinfo{volume}{605}},
  \bibinfo{pages}{1611 } (\bibinfo{year}{2011}).

\bibitem[{\citenamefont{Cai et~al.}(2014)\citenamefont{Cai, Lan, Zhang, and
  Zhang}}]{PhysRevB.89.035438}
\bibinfo{author}{\bibfnamefont{Y.}~\bibnamefont{Cai}},
  \bibinfo{author}{\bibfnamefont{J.}~\bibnamefont{Lan}},
  \bibinfo{author}{\bibfnamefont{G.}~\bibnamefont{Zhang}}, \bibnamefont{and}
  \bibinfo{author}{\bibfnamefont{Y.-W.} \bibnamefont{Zhang}},
  \bibinfo{journal}{Phys. Rev. B} \textbf{\bibinfo{volume}{89}},
  \bibinfo{pages}{035438} (\bibinfo{year}{2014}).

\bibitem[{\citenamefont{Molina-S\'anchez and Wirtz}(2011)}]{PhysRevB.84.155413}
\bibinfo{author}{\bibfnamefont{A.}~\bibnamefont{Molina-S\'anchez}}
  \bibnamefont{and} \bibinfo{author}{\bibfnamefont{L.}~\bibnamefont{Wirtz}},
  \bibinfo{journal}{Phys. Rev. B} \textbf{\bibinfo{volume}{84}},
  \bibinfo{pages}{155413} (\bibinfo{year}{2011}).

\end{thebibliography}

\end{document}